\documentclass[12pt]{article}
\pdfoutput=1
\usepackage{tikz}
\usepackage{putex}
\usepackage[vcentermath]{youngtab}
\usepackage{subfig}
\usepackage{lscape}

\usepackage{graphicx}
\usepackage{epstopdf}
\usepackage{enumerate}
\usepackage{cite}
\usepackage{tensor} 
\usepackage{slashed}
\usepackage{amsmath}
\usepackage{amssymb}
\usepackage{mathrsfs}
\usepackage{lgrind}
\usepackage{youngtab}

\usepackage{amsmath,amssymb,braket}

\usepackage{bbm}

\usepackage{hyperref}

\numberwithin{equation}{section}

\newcommand {\be} {\begin {equation}}
\newcommand {\ee} {\end {equation}}

\newcommand {\bes} {\begin {equation*}}
\newcommand {\ees} {\end {equation*}}

\def\CH{{\cal H}}

\newcommand{\beq}{\begin{equation}}
\newcommand{\eeq}{\end{equation}}

\def\be{ \begin{equation} }
\def\ee{ \end{equation} }

\def \be {\beta}

\def \beq { \begin{equation}}
\def \eeq {\end{equation}}

\begin{document}

\preprint{PUPT-2568}

\institution{PU}{Department of Physics, Princeton University, Princeton, NJ 08544, USA}
\institution{PCTS}{Princeton Center for Theoretical Science, Princeton University, Princeton, NJ 08544, USA}
\institution{DEI}{Department of Physics and Computer Science, Dayalbagh Educational Institute, Agra 282005, India}
\institution{HU}{Department of Physics, Harvard University, Cambridge, MA 02138, USA}

\title{
Prismatic Large $N$ Models for Bosonic Tensors
}

\authors{Simone Giombi\worksat{\PU}, Igor R.~Klebanov\worksat{\PU,\PCTS}, Fedor Popov\worksat{\PU}, \\[10pt]
Shiroman Prakash\worksat{\DEI}, Grigory Tarnopolsky\worksat{\HU}
}

\abstract{We study the $O(N)^3$ symmetric quantum field theory of a bosonic tensor $\phi^{abc}$ with sextic interactions. Its large $N$ limit is dominated 
by a positive-definite operator, whose index structure has the topology of a prism. We present a large $N$ solution of the model using Schwinger-Dyson equations to sum
the leading diagrams, 
finding that for $2.81 < d < 3$ and for $d<1.68$ the spectrum of bilinear operators has no complex scaling dimensions. 
We also develop perturbation theory 
in $3-\epsilon$ dimensions including eight $O(N)^3$ invariant operators necessary for the renormalizability. For sufficiently large $N$, 
we find a ``prismatic" fixed point of the renormalization group, where all eight coupling constants are
real. The large $N$ limit of the resulting $\epsilon$ expansions of various operator dimensions agrees with the Schwinger-Dyson equations. 
Furthermore, the $\epsilon$ expansion allows us to calculate the $1/N$ corrections to operator dimensions.
The prismatic fixed point in $3-\epsilon$ dimensions survives down to $N\approx 53.65$, where it merges with another fixed point and becomes complex.
We also discuss the $d=1$ model where our approach gives a slightly negative scaling dimension for $\phi$, while the spectrum of bilinear operators
is free of complex dimensions.
}

\date{}

\maketitle

\tableofcontents

\section{Introduction}

In recent literature, there has been considerable interest in models where the degrees of freedom transform as tensors of rank $3$ or higher.
Such models with appropriately chosen interactions admit new kinds of large $N$ limits, which are not of 't Hooft type
 and are dominated by the so-called melonic Feynman diagrams \cite{Gurau:2009tw,Bonzom:2011zz,Carrozza:2015adg,Witten:2016iux, Klebanov:2016xxf}. 
Much of the recent activity (for a review see \cite{Klebanov:2018fzb}) has been on the quantum mechanical models of fermionic tensors \cite{Witten:2016iux, Klebanov:2016xxf},
which have 
large $N$ limits similar to that in the Sachdev-Ye-Kitaev (SYK) model \cite{Sachdev:1992fk,Kitaev:2015,Polchinski:2016xgd,Maldacena:2016hyu,Jevicki:2016bwu,
Gross:2016kjj,Kitaev:2017awl,Rosenhaus:2018dtp}.

It is also of interest to explore similar quantum theories of bosonic tensors \cite{Klebanov:2016xxf,Giombi:2017dtl,Murugan:2017eto}.
In \cite{Klebanov:2016xxf,Giombi:2017dtl} an $O(N)^3$ invariant theory of the scalar fields $\phi^{abc}$ was studied:
\begin{align}
S_4= \int d^d x \left ( \frac {1} {2} (\partial_\mu \phi^{abc})^2 + 
{g \over 4!} O_{\rm tetra}
\right )\ , \notag \\
O_{\rm tetra}=  \phi^{a_1 b_1 c_1} \phi^{a_1 b_2 c_2} \phi^{a_2 b_1 c_2} \phi^{a_2 b_2 c_1}
\ .
\label{tetra}
\end{align}
This QFT is super-renormalizable in $d<4$
and is formally solvable using the Schwinger-Dyson equations in the large $N$ limit where $g N^{3/2}$ is held fixed. However, this model has some instabilities. 
One problem is that the ``tetrahedral" operator $O_{\rm tetra}$ is not positive definite. Even if we ignore this and consider the large $N$ limit formally, we find that in $d<4$ the 
$O(N)^3$ invariant operator $\phi^{abc} \phi^{abc}$ has a complex dimension of the form $\frac{d}{2}+ i\alpha(d)$ \cite{Giombi:2017dtl}.\footnote{Such complex dimensions
appear in various other large $N$ theories; see, for example, \cite{Dymarsky:2005uh,Pomoni:2008de,Grabner:2017pgm,Prakash:2017hwq}.}
From the dual AdS point of view, such a complex dimension corresponds to a scalar field whose $m^2$ is below the 
Breitenlohner-Freedman stability bound \cite{Breitenlohner:1982jf,Klebanov:1999tb}. 
The origin of the complex dimensions was elucidated using perturbation theory in $4-\epsilon$ dimensions: the fixed point was found to be at 
complex values of the couplings for the additional $O(N)^3$ invariant operators required by the renormalizability \cite{Giombi:2017dtl}. 
In \cite{Giombi:2017dtl} a $O(N)^5$ symmetric theory for tensor $\phi^{abcde}$ and sextic interactions was also considered. It was found that the
dimension of operator $\phi^{abcde}\phi^{abcde}$ is real in the narrow range $d_{\rm crit}< d< 3$, where $d_{\rm crit}\approx 2.97$. However, the scalar potential of this theory is
again unstable, so the theory may be defined only formally. 
 In spite of these problems, some interesting formal results on melonic scalar theories of this type were found recently \cite{Liu:2018jhs}. 

\begin{figure}[h!]
\begin{center}
\includegraphics[width=15.5cm]{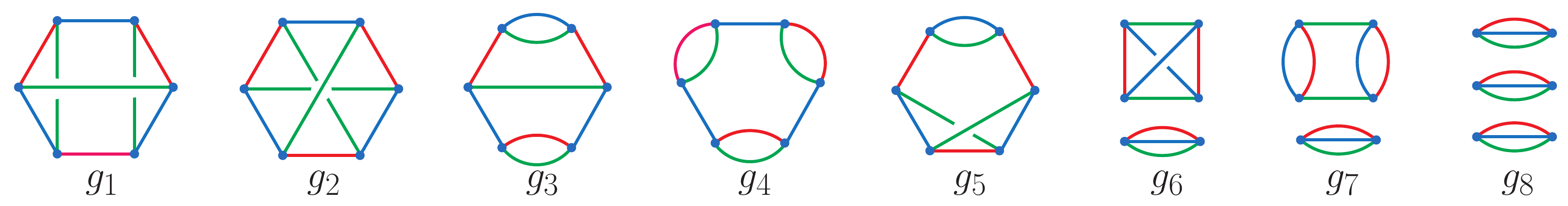}
\caption{Diagrammatic representation of the eight possible $O(N)^3$ invariant sextic interaction terms. }
\label{8inter}
\end{center}
\end{figure}  

In this paper, we continue the search for stable bosonic large $N$ tensor models with multiple $O(N)$ symmetry groups. 
Specifically, we study the $O(N)^3$ symmetric theory of scalar fields $\phi^{abc}$ with a sixth-order interaction, whose Euclidean action is
\begin{equation}
S_6= \int d^d x \left ( \frac {1} {2} (\partial_\mu \phi^{abc})^2 + 
{g_1\over 6!} \phi^{a_1 b_1 c_1} \phi^{a_1 b_2 c_2} \phi^{a_2 b_1 c_2} \phi^{a_3 b_3 c_1} \phi^{a_3 b_2 c_3} \phi^{a_2 b_3 c_3}
\right )
\ .
\label{prism}
\end{equation}
This QFT is super-renormalizable in $d<3$.
When the fields $\phi^{abc}$ are represented by vertices and index contractions by edges, this interaction term looks like a prism 
(see figure 11 in \cite{Klebanov:2016xxf});
it is the leftmost diagram in figure \ref{8inter}.
Unlike with the tetrahedral quartic interaction (\ref{tetra}), the action (\ref{prism}) is positive for $g_1>0$.  
In sections \ref{LargeN} and \ref{bilinears}, 
we will show that there is a smooth large $N$ limit where $g_1 N^{3}$ is held fixed and derive formulae for various operator dimensions in continuous $d$.
We will call this large $N$ limit the ``prismatic" limit: the leading Feynman diagrams are not the same as the melonic diagrams, which appear
in the $O(N)^5$ symmetric $\phi^6$ QFT for a tensor $\phi^{abcde}$ \cite{Giombi:2017dtl}. However, as we discuss in section \ref{LargeN}, 
the prismatic interaction may be reduced to
a tetrahedral one, (\ref{prismaux}), by introducing
an auxiliary tensor field $\chi^{abc}$.

The theory (\ref{prism}) may be viewed as a tensor counterpart of the bosonic theory with random couplings, which was introduced in section 6.2 of
\cite{Murugan:2017eto}. Since both theories are dominated by the same class of diagrams in the large $N$ limit, they have the same Schwinger-Dyson equations for the 2-point and
4-point functions. We will confirm the conclusion of \cite{Murugan:2017eto} that the $d=2$ theory does not have a stable IR limit; this is due to the
appearance of a complex scaling dimension. However, we find that in the 
ranges $2.81 < d < 3$ and $d<1.68$, the large $N$ prismatic theory 
does not have  any complex dimensions for the bilinear operators.
In section \ref{threeminuseps} we use renormalized perturbation theory to develop the $3-\epsilon$ expansion of the prismatic QFT. 
We have to include all eight operators invariant under the
$O(N)^3$ symmetry and the $S_3$ symmetry permuting the $O(N)$ groups; they are shown in figure \ref{8inter} and written down in (\ref{allinter}).
For $N> N_{\rm crit}$, where  $N_{\rm crit}\approx 53.65$, 
we find a prismatic RG fixed point where all eight coupling constants are
real. At this fixed point, $\epsilon$ expansions of various operator dimensions agree in the large $N$ limit with those obtained using the Schwinger-Dyson equations. 
Futhermore, the $3-\epsilon$ expansion provides us with a method to calculate the $1/N$ corrections to operator dimensions, as shown in
(\ref{deltaphi}), (\ref{deltaphisquared}).
At $N = N_{\rm crit}$ the prismatic fixed point merges with another fixed point, and for $N < N_{\rm crit}$ both become complex.

In section \ref{BosonicQM} we discuss the $d=1$ version of the model (\ref{prism}).
Our large $N$ solution gives a slightly negative scaling dimension, $\Delta_\phi\approx -0.09$, while
the spectrum of bilinear operators
is free of complex scaling dimensions.

\section{Large $N$ Limit}

\label{LargeN}

To study the large $N$ limit of this theory,
we will find it helpful to introduce an auxiliary field $\chi^{abc}$ so that\footnote{If we added fermions to make the tensor model 
supersymmetric \cite{Klebanov:2016xxf, Murugan:2017eto, Bulycheva:2018qcp, Chang:2018sve}
then $\chi^{abc}$ would be interpreted as the highest component of the superfield $\Phi^{abc}$.}
\begin{equation}
S= \int d^d x \left ( \frac {1} {2} (\partial_\mu \phi^{abc})^2 + 
{g\over 3!} \phi^{a_1 b_1 c_1} \phi^{a_1 b_2 c_2} \phi^{a_2 b_1 c_2} \chi^{a_2 b_2 c_1}-\\\frac{1}{2}\chi^{abc}\chi^{abc}\right )\ .
\label{prismaux}
\end{equation}
where $g\sim \sqrt g_1$.
Integrating out $\chi^{abc}$ gives rise to the action (\ref{prism}). The advantage of keeping $\chi^{abc}$ explicitly is that the theory is then
a theory with $O(N)^3$ symmetry dominated by the tetrahedral interactions, except it now involves two  rank-3 fields; this shows that it has a smooth large $N$ limit. 
Thus, a prismatic large $N$ limit for the theory with one 3-tensor $\phi^{abc}$ may be viewed as a tetrahedral limit for two 3-tensors. 

\begin{figure}[h]
\begin{center}
\includegraphics[width=10cm]{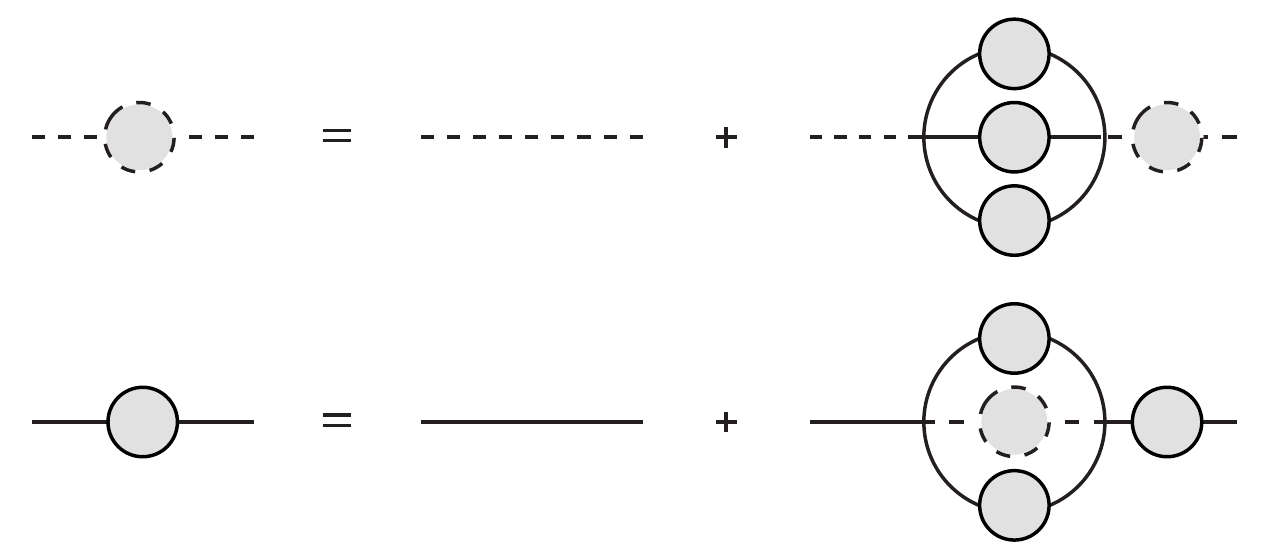}
\caption{Diagramatic representation of the Schwinger-Dyson equations. Solid lines denote $\phi$ propagators, and dashed lines denote $\chi$ propagators. \label{SD}}
\end{center}
\end{figure}

Let us define the following propagators:
\begin{align}
\langle \phi(p)\phi(q) \rangle  = (2\pi)^d \delta^d(p+q) G(p), \quad \langle \chi(p)\chi(q) \rangle =(2\pi)^d\delta^d(p+q) F(p).
\end{align}
In the free theory $G(p)=G_0(p)=\frac{1}{p^2}$, and $F(p)=F_0=1$.
In the strong coupling limit the self-energies of the fields are given by the inverse propagators: $G(p)^{-1} = \Sigma_\phi$ and $F(p)^{-1}=\Sigma_\chi$.
The Schwinger-Dyson equations for the exact two-point functions can be written as:
\begin{eqnarray}
F(p) & = & F_0 + g^2N^3 F_0 \int \frac{d^dq d^d k}{(2\pi)^{2d}} G(p-q-k)G(q)G(k) F(p)\,, \notag \\
G(p) & = & G_0(p) + 3 g^2N^3 G_0(p) \int \frac{d^dq d^d k}{(2\pi)^{2d}} G(p-q-k)F(q)G(k) G(p)\,, 
\end{eqnarray}
and represented in figure \ref{SD}.

Multiplying the first equation by $F_0^{-1}$ on the left and $F(p)^{-1}$ on the right, and likewise for the second equation we obtain: 
\begin{eqnarray}
F(p)^{-1} & = & F_0^{-1} - \lambda^2 \int \frac{d^dq d^d k}{(2\pi)^{2d}} G(p-q-k)G(q)G(k)\,,  \notag \\
G(p)^{-1} & = & G_0(p)^{-1} - 3 \lambda^2  \int \frac{d^dq d^d k}{(2\pi)^{2d}} G(p-q-k)F(q)G(k)\,,
\end{eqnarray}
where $\lambda^2=N^3g^2\sim N^3 g_1$. We have to take the large $N$ limit keeping $\lambda^2$ fixed.
In the IR limit, let us assume $$G(p)=\frac{A}{p^{2a}},~F(p)=\frac{B}{p^{2b}}.$$ $a$ is related to the scaling dimension of $\phi$, $\Delta_\phi$ via $a=d/2-\Delta_\phi.$ 

For what range of $a$ and $b$ can we drop the free terms in the gap equations above? In the strong coupling limit we require $b<0$ and $a<1$.
Since
 $b=-3a+d$, we have
$
d/3<a<1
$. 
In terms of $\Delta_\phi$, we then find
\begin{equation}
3 \Delta_\phi+ \Delta_\chi=d\ , \qquad d/2-1<\Delta_\phi<d/6.
\label{allowedrange}
\end{equation} 
Notice that, if we had the usual kinetic term for the $\chi$ field, the allowed range for $\Delta_\phi$ would be larger. 
Therefore, our solution may also apply to a model with two dynamical scalar fields interacting via the particular interaction given above. 

The gap equation is finally:
\begin{eqnarray}
F(p)^{-1} & = &  - \lambda^2 \int \frac{d^dq d^d k}{(2\pi)^{2d}} G(p-q-k)G(q)G(k)\,, \notag \\
G(p)^{-1} & = &  - 3 \lambda^2  \int \frac{d^dq d^d k}{(2\pi)^{2d}} G(p-q-k)F(q)G(k)\,. 
\end{eqnarray}
Dimensional analysis of the strong coupling fixed point actually does not fix $a$: we get $b=-3a+d$ from the first equation and $a = -2a-b+d$ from the second equation.
Let us 
try to solve the above equations, in the hope that numerical factors arising from the Feynman integrals may determine $a$. The overall constant $A$ is not determined from this procedure, but note that $[\lambda]=3-d$, and therefore $A \sim \lambda^{\frac{2(a-1)}{3-d}}$. This procedure is analogous to solving an eigenvalue equation, and perhaps it is not surprising that we have to do this, since a solution for $a$ also determines the anomalous dimension of a composite operator $\phi^3$.
We then find
\begin{equation}
F(p)=\frac{-1}{A^3\lambda^2} \frac{(2\pi)^{2d}}{L_d(a,a)L_d(2a-d/2,a)} \frac{1}{p^{2b}}, 
\end{equation}
where 
\begin{equation}
L_{d}(a,b) = \pi^{d/2}\frac{\Gamma(d/2-a)\Gamma(d/2-b)\Gamma(a+b-d/2)}{\Gamma(a)\Gamma(b)\Gamma(d-a-b)}. 
\end{equation}

The condition that must be satisfied by $a$ is then:
\begin{equation}
3 \frac{L_{d}(2a-d/2,d-3a)}{L_{d}(2a-d/2,a)}=1\ . \label{eigenvalue}
\end{equation}

In position space, the IR two-point functions take the form
\begin{eqnarray}
G(x) & = & \frac{\Gamma(d/2-a)}{\pi^{d/2}2^{2a}\Gamma(a)} \frac{A}{(x^2)^{\Delta_\phi}},\\
F(x) & = & \frac{\Gamma(d/2-b)}{\pi^{d/2}2^{2b}\Gamma(b)}  \frac{(2\pi)^{2d}}{A^3\lambda^2L_d(a,a)L_d(2a-d/2,a)} \frac{1}{(x^2)^{d-3\Delta_\phi}}.
\end{eqnarray}

In terms of $\Delta_\phi$, (\ref{eigenvalue}) may be written as
\begin{equation}
f(d,\Delta_\phi) \equiv \frac{1}{3}{\Gamma (\frac{d}{2} - 3 \Delta_\phi)  \Gamma (-\frac{d}{2} + 3\Delta_\phi) \Gamma (\Delta_\phi) \Gamma (d-\Delta_\phi)\over \Gamma (\frac{d}{2} - \Delta_\phi) 
\Gamma (-\frac{d}{2} + \Delta_\phi) \Gamma (3 \Delta_\phi) \Gamma (d- 3\Delta_\phi)}
 =1 \,. \label{eigenvaluenew}
\end{equation}

\subsection{The scaling dimension of $\phi$}

\begin{figure}[h]
\begin{center}
\includegraphics[width=10cm]{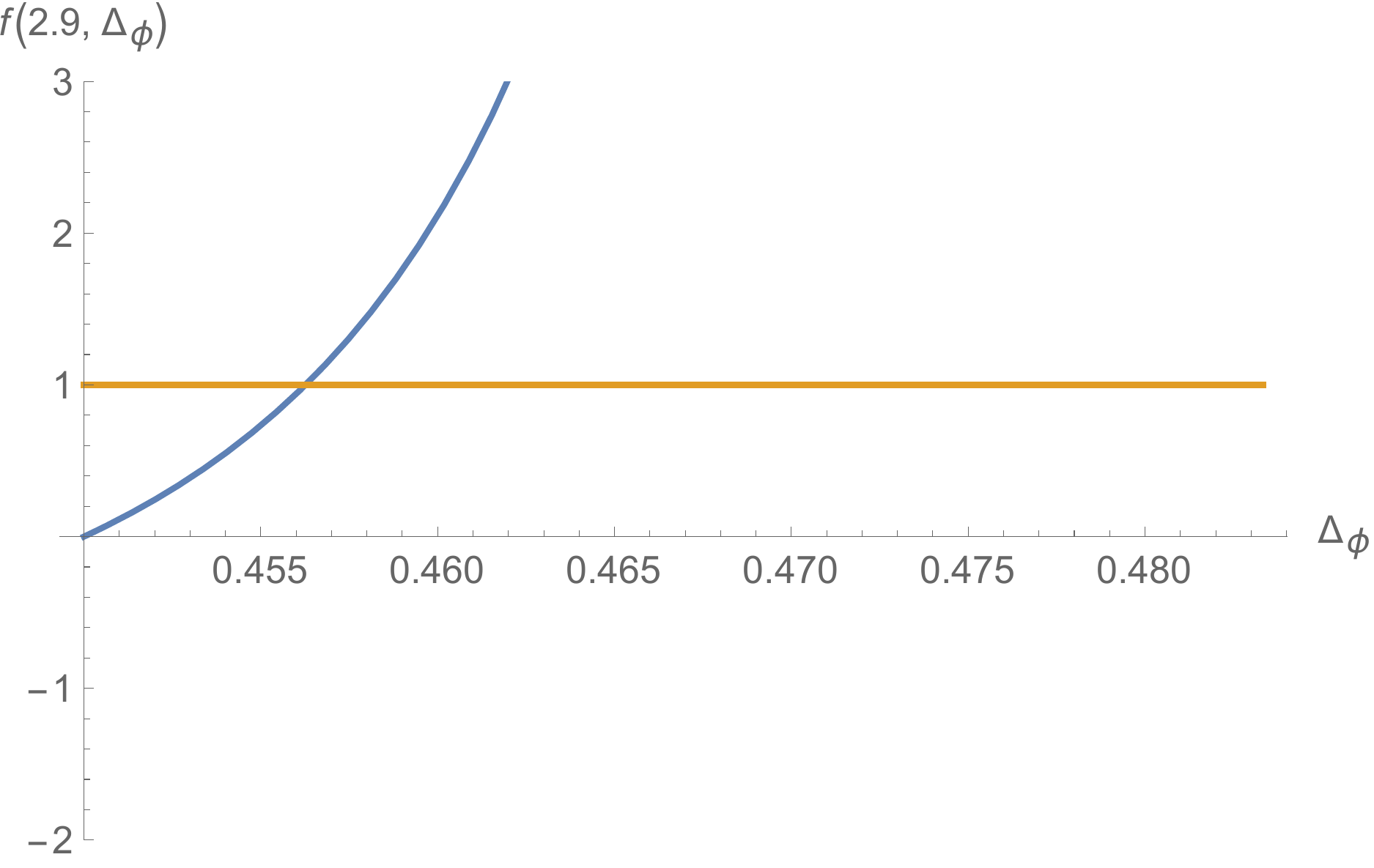}
\caption{Solving \eqref{eigenvaluenew} for $d=2.9$. 
\label{2.9d}}
\end{center}
\end{figure}

It can be verified numerically that that solutions to \eqref{eigenvaluenew} within the allowed range (\ref{allowedrange}) do exist in $d<3$.
For example, for $d=2.9$ we have the solution shown in Figure \ref{2.9d}:
\begin{align}
\Delta_\phi \approx 0.456\ ,\qquad
\Delta_\chi \approx 1.531\ .
\end{align}
For $d=2.99$, we find $\Delta_\phi=0.495$, and $d=2.999$, $\Delta_\phi=0.4995$, consistent with 
the $3-\epsilon$ expansion (\ref{SDdim}).
For $d=2$, \eqref{eigenvalue} simplifies to
\begin{equation}
3 (3\Delta_\phi-1)^2 = (\Delta_\phi-1)^2\ .
\end{equation}
The solution $\Delta_\phi = \frac{1}{13} \left(4-\sqrt{3}\right)$ lies within the allowed range (\ref{allowedrange}), while 
the one with the other branch of the square root is outside it.

\begin{figure}[h!]
\begin{center}
\includegraphics[width=10cm]{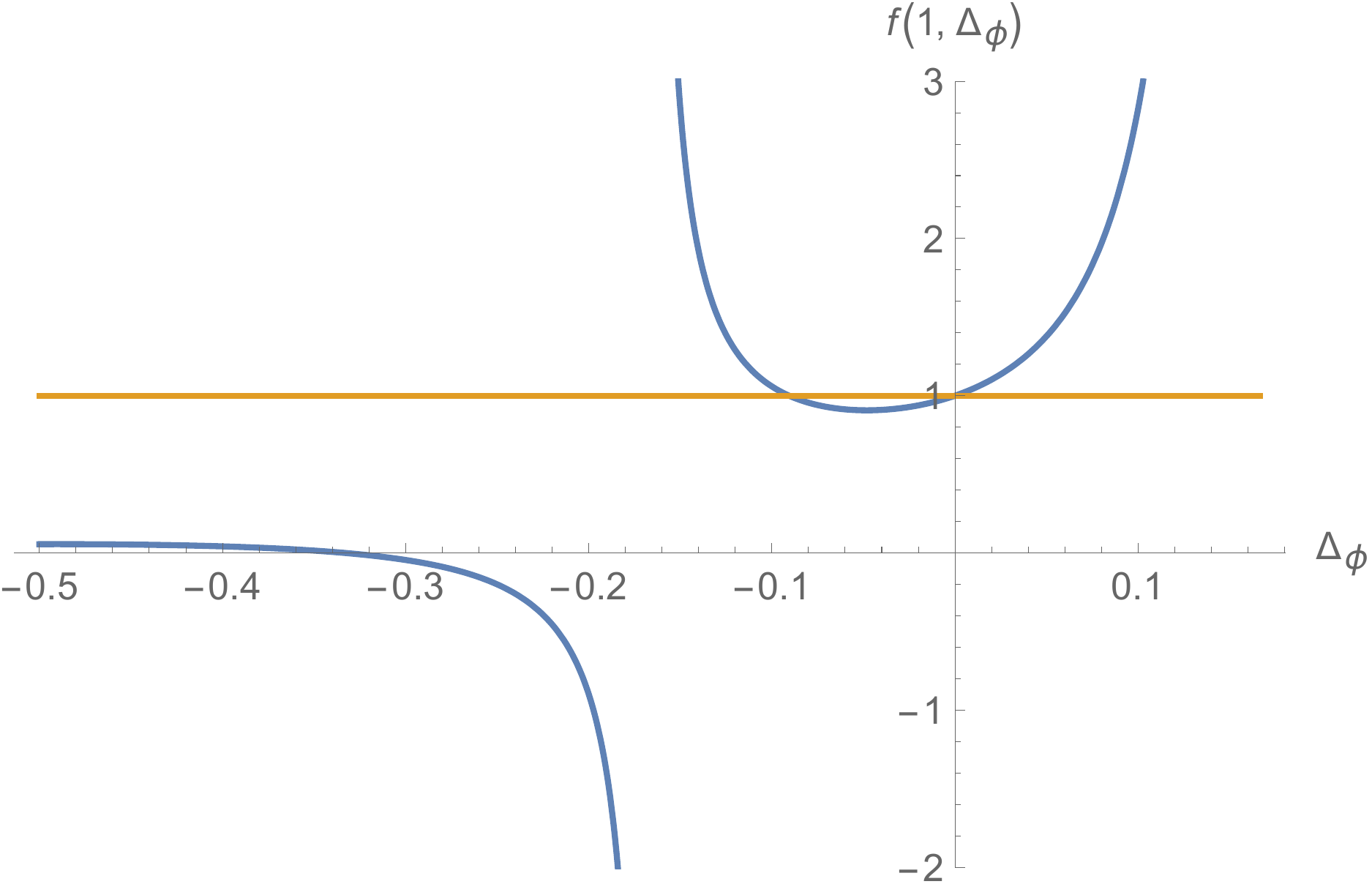}
\caption{Solving \eqref{eigenvaluenew} for $d=1$. 
\label{1d}}
\end{center}
\end{figure}

For $d<2$ we find multiple solutions within the allowed range (\ref{allowedrange}), as shown for $d=1$ in figure \ref{1d}. 
One of the solutions gives $\Delta_\phi=0$; this produces singularities in the large $N$ dimensions of scalar bilinears, and we will not use it.
The other solution, 
\begin{equation}
\Delta_\phi\approx -0.09055\ , \qquad 
\Delta_\chi \approx 1.2717\ ,
\label{QMsol}
\end{equation}
appears to be acceptable. Although $\Delta_\phi$ is negative, it lies above the unitarity bound.
We note that there is also a positive solution
$\Delta_\phi \approx 0.225$, which lies outside of the allowed range (although it would be allowed if the $\chi$ field was dynamical).

There is an interesting transition in behavior which happens at $d=d_c$ where there is a double root at $\Delta_\phi=0$.
The critical dimension $d_c$ is the solution of
\begin{equation}
2+d_c \pi \cot(d_c \pi/2)+d_c (\gamma+\psi(d_c)) = 0\ .
\end{equation}
Its numerical value is $d_c = 1.35287$.
For $d$ slightly above $d_c$ one of the solutions for $\Delta_\phi$ is zero, while the other is positive; we have to pick the positive one.
However, for $d$ slightly below $d_c$ one of the solutions for $\Delta_\phi$ is zero, while the other is negative.
Special care may be needed for continuation to $d< d_c$; in particular, for studying the $d=1$ case.

\section{Bilinear Operators}
\label{bilinears}

There are three types of scalar bilinears one can consider, which are of the schematic form: $A=\phi (\xi \cdot \partial)^s (\partial^{2})^n  \phi$, $B=\phi  (\xi \cdot \partial)^s (\partial^{2})^n \chi$ 
and $C=\chi  (\xi \cdot \partial)^s (\partial^{2})^n \chi$, where $\xi^\mu$ is an auxiliary null vector introduced to encode the spin of the operators,
$ \xi \cdot \partial= \xi^\mu \partial_\mu$, and $\partial^2= \partial^\mu \partial_\mu$.
We note that there is mixing of operators of type $A$ and $C$. It is easy to convince oneself that there is no mixing with the $B$ operators by drawing a few diagrams.

\subsection{Bilinears of type B}

\begin{figure}
\begin{center}
\includegraphics[width=3cm]{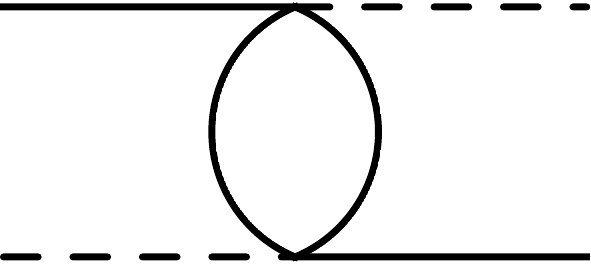}
\caption{The integration kernel for type B bilinears. \label{kernelB}}
\end{center}
\end{figure}

Let us consider a bilinear of type $B$, of spin $s$ and scaling dimension $\Delta$, for which there is no mixing. The three-point functions take the form \cite{Osborn:1993cr,Giombi:2011rz}:
\begin{equation}
\begin{split}
\langle  \phi^{abc}(x_1)\chi^{abc}(x_2) B_s (x_3;\xi)\rangle & = v^{(B)}(x_1,x_2,x_3) = \frac{Q_3^s}{x_{31}^{\tau+\Delta_\phi-\Delta_\chi} x_{32}^{\tau+\Delta_\chi-\Delta_\phi}x_{12}^{\Delta_\phi+\Delta_\chi-\tau}} \\
&  \rightarrow v^{(B)}_{s,\tau}(x_1,x_2) = (x_{12}\cdot \xi)^s x_{12}^{\tau-\Delta_\phi-\Delta_\chi}\,,
\end{split}
\end{equation}
where $\tau=\Delta-s$ is the twist of the bilinear, $\xi$ is the null polarization vector, $Q_3$ is the conformally invariant tensor structure defined in \cite{Osborn:1993cr,Giombi:2011rz} and we took the limit ${x_3\rightarrow \infty}$ in the second line. 
The eigenvalue equation, obtained using the integration kernel depicted schematically in figure \ref{kernelB}, is
\begin{equation}
v_{s,\tau}(x_1,x_2) = 3\lambda^2 \int d^d y d^d z  F(x_2,y)G(y,z)^2G(z,x_1)v_{s,\tau}(y,z)
\end{equation}

When $s=0$, we have:
\begin{equation}
|x_1-x_2|^{-\Delta_\phi-\Delta_\chi+\Delta}=3\tilde{A}^3 \tilde{B}\lambda^2  \int d^d y d^d z \frac{1}{|x_2-y|^{2\Delta_\chi}|y-z|^{5\Delta_\phi+\Delta_\chi-\Delta}|z-x_1|^{2\Delta_\phi}}
\end{equation}
which translates into 
\begin{eqnarray}
\!\!\!\!\!\!\!\!\!\!\!\!\!\!
g^{(B)}(d,\Delta) &\equiv& -3\frac{\Gamma (3 \Delta_\phi ) \sin
   \left(\frac{1}{2} \pi  (d-6
   \Delta_\phi )\right) \Gamma
   \left(\frac{d}{2}-\Delta_\phi \right)
   \Gamma \left(-\frac{d}{2}+3
   \Delta_\phi +1\right) \Gamma
   \left(\frac{\Delta}{2}-\Delta_\phi \right)
   \Gamma \left(\frac{1}{2} (d-\Delta-2
   \Delta_\phi )\right)}{\pi  \Gamma
   (\Delta_\phi ) \Gamma
   \left(\frac{\Delta}{2}+\Delta_\phi \right)
   \Gamma \left(\frac{d-\Delta}{2}+\Delta_\phi
   \right)}\cr
&=&1 \ .
\label{typeBscalars}
\end{eqnarray}

\begin{figure}[h!]
\begin{center}
\includegraphics[width=10cm]{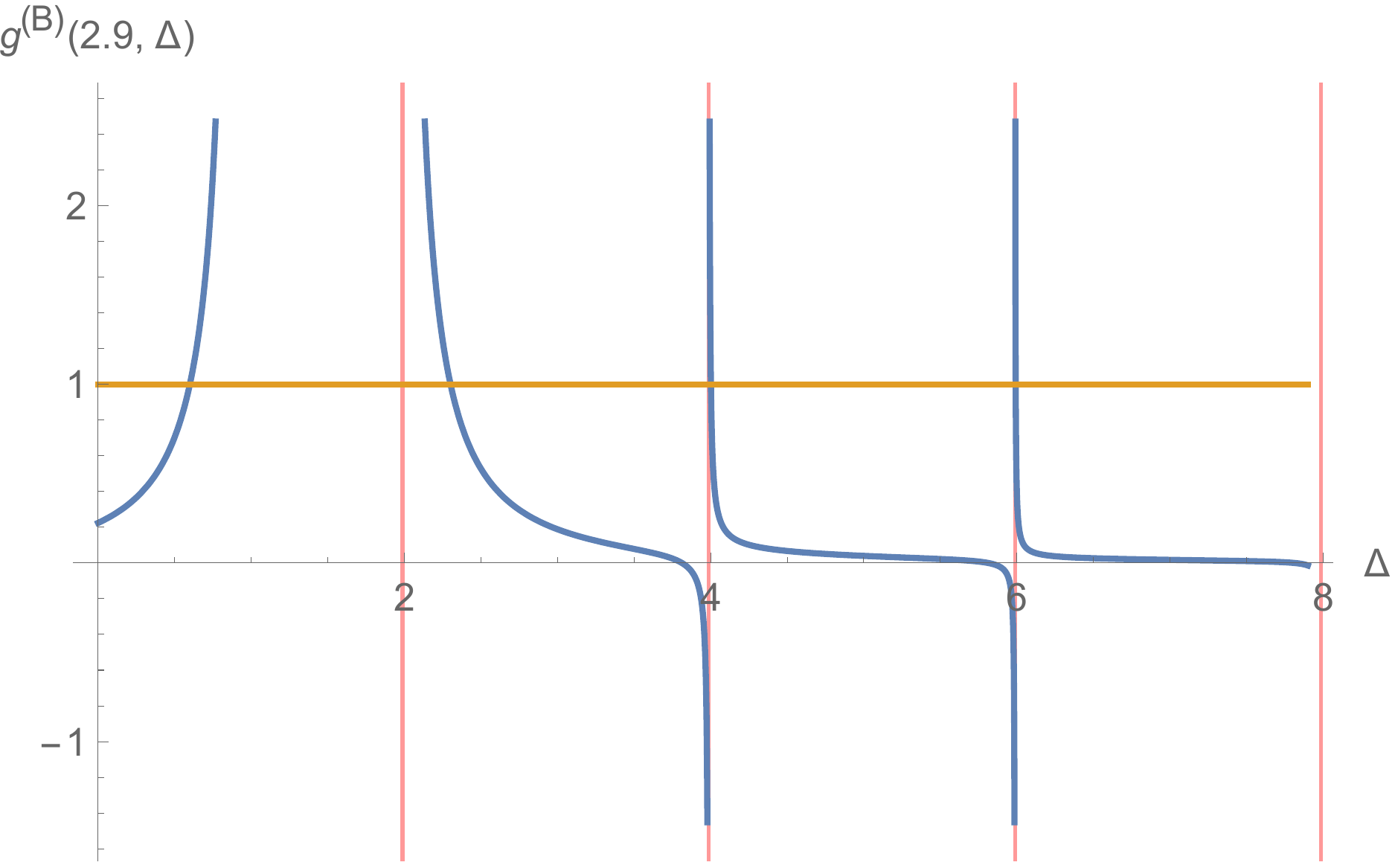}
\caption{The spectrum of type B bilinears in $d=2.9$. The red lines correspond to asympotes at $2n+\Delta_\phi+\Delta_\chi=2n+1.98747$. \label{typeB} }
\end{center}
\end{figure}

We can solve equation (\ref{typeBscalars}) numerically to find the allowed scaling dimensions for type B operators in various dimensions. In $d=2.9$ the type B scaling dimensions are 
\begin{equation}
\Delta_B=2.30120;~4.00173;~5.99214;~7.98983; ~9.98891;~\ldots,
\end{equation} 
as shown in figure \ref{typeB}. In the pure $\phi$ language, the first one can be identified with the
tetrahedral operator. 
The type B scaling
 approach the asymptotic formula
\begin{align}
\Delta_B\rightarrow 2n+\Delta_\phi+ \Delta_\chi= 2n+ 1.98747\ .
\label{asympd29}
\end{align} 
For example, for $n=54$ we numerically find $\Delta= 109.98749$, which is very close to (\ref{asympd29}).

For spin $s>0$ the eigenvalue equation is:
\begin{equation}
(x_{12}\cdot \xi)^s|x_1-x_2|^{-\Delta_\phi-\Delta_\chi+\Delta}=3\tilde{A}^3 \tilde{B}\lambda^2  \int d^d y d^d z \frac{((y-z)\cdot \xi)^s}{|x_2-y|^{2\Delta_\chi}|y-z|^{5\Delta_\phi+\Delta_\chi-\Delta}|z-x_1|^{2\Delta_\phi}}
\end{equation}
Note that the spectrum of type $B$ bilinears does not contain the stress tensor, which is of type $A/C$.

\begin{figure}[h]
\begin{center}
\includegraphics[width=10cm]{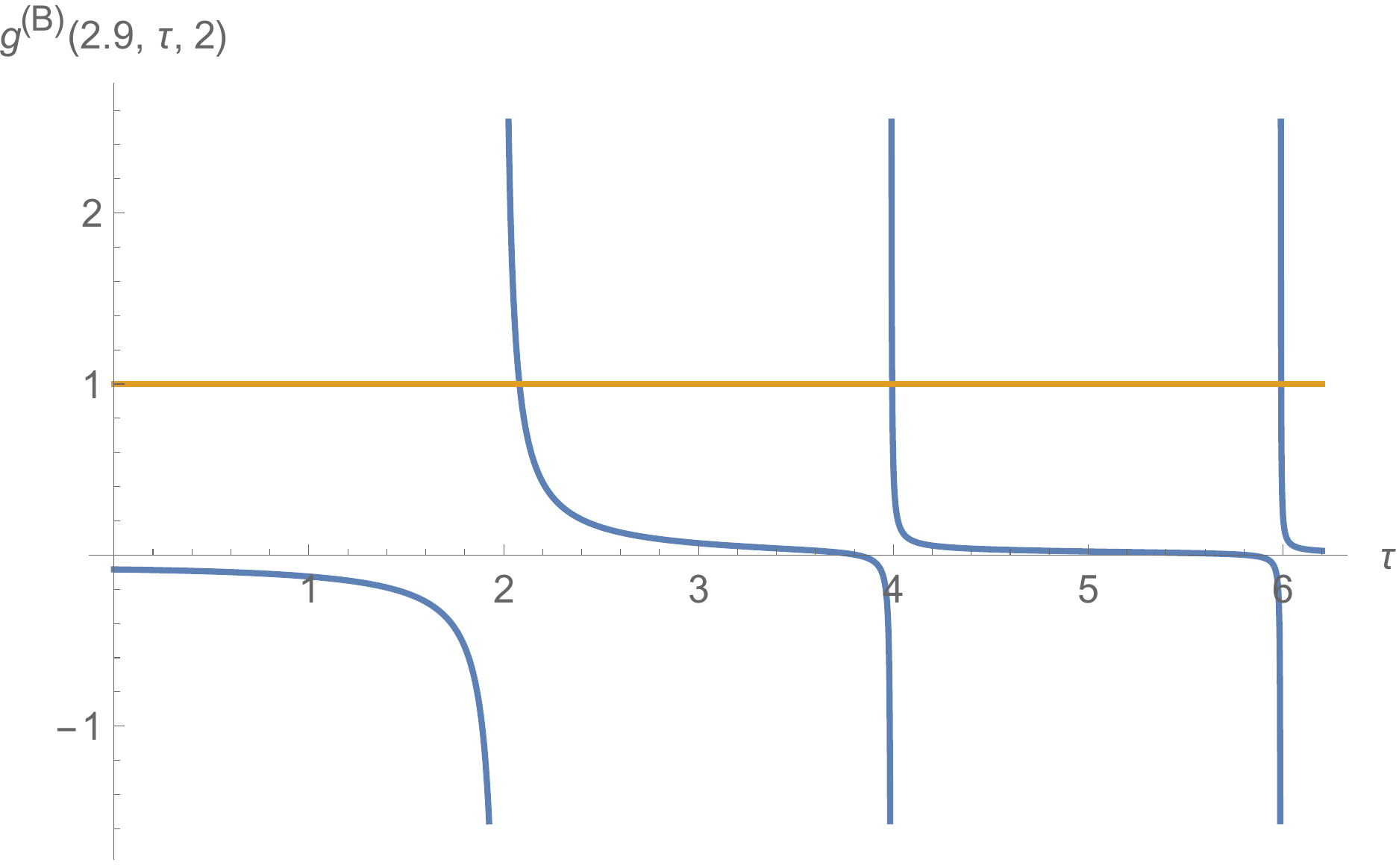}
\caption{Solving equation \eqref{hsBeq} in $d=2.9$ for the allowed twists of spin-$2$ type $B$ bilinears. \label{hsB}}
\end{center}
\end{figure}

Processing the equation we have the following condition for the allowed twists of higher spin bilinears:
\begin{eqnarray}
\!\!\!\!\!\!\!\!\!\!\!\!\!\!
 && g^{(B)}(d,\tau,s) \equiv \nonumber \\
 & & -3\frac{\Gamma (3 \Delta_\phi) \sin
   \left(\frac{1}{2} \pi  (d-6
   \Delta_\phi)\right) \Gamma
   \left(\frac{d}{2}-\Delta_\phi\right)
   \Gamma \left(-\frac{d}{2}+3
   \Delta_\phi+1\right) \Gamma
   \left(\frac{1}{2} (d-2 \Delta_\phi
   -\tau )\right) \Gamma
   \left(s-\Delta_\phi+\frac{\tau
   }{2}\right)}{\pi  \Gamma (\Delta_\phi
   ) \Gamma \left(\frac{d}{2}+\Delta_\phi
   -\frac{\tau }{2}\right) \Gamma
   \left(s+\Delta_\phi+\frac{\tau
   }{2}\right)} \nonumber \\
   & & =  1 \ .\label{hsBeq}\ 
\end{eqnarray}
Using this equation one can find the allowed twists of spin-$s$ type B bilinears. For example,
the spectrum when $s=2$ and $d=2.9$ is found from figure \ref{hsB} to be $\tau=2.08,~3.99,~5.99,~7.99, \ldots$, which approach $\Delta_\chi+\Delta_\phi
+2n=1.99+2n$ from above.

We find that the spectrum of type B bilinear appears to be real for all $d<3$. However, 
there are ranges of $d$ where
the spectrum of type A/C operators do contain complex eigenvalues, as we discuss in the next section.

\subsection{Mixing of bilinears of type A and C}

\begin{figure}
\begin{center}
\includegraphics[width=3cm]{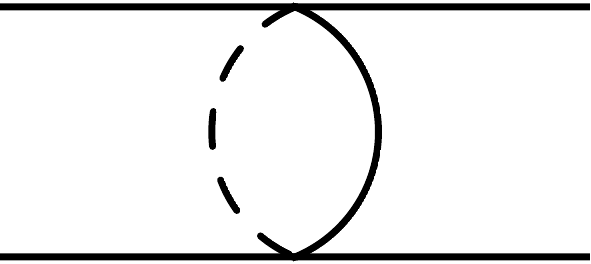} $~~~$ \includegraphics[width=3cm]{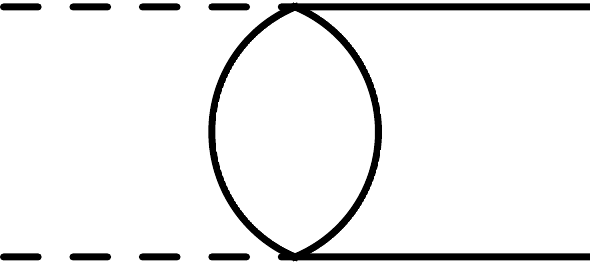}  
$~~~$ \includegraphics[width=3.1cm]{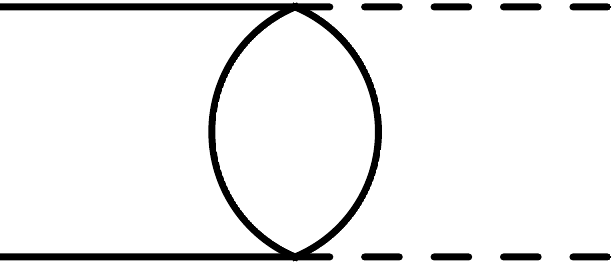}  
\caption{The integration kernels $K_{AA}$, $K_{CA}$ and $K_{AC}$ respectively for mixtures of type $A$ and $C$ bilinears. \label{kernelA}}
\end{center}
\end{figure}

Let us now study the spectrum of bilinear operators of type $A$ and $C$. As mentioned earlier, by drawing a few diagrams (see figure \ref{kernelA}) one can see that these operators mix, in the sense that the two-point function $\langle A_s C_s \rangle \neq 0$.  Let $\tau=\Delta-s$ be the twist of mixture of $A$ and $C$ operators, which we denote as $\tilde{A}_s$. As in the previous subsection, from the three-point functions $\langle  \phi^{abc}(x_1)\phi^{abc}(x_2) \tilde{A}_s (x_3;\xi)\rangle $ and $\langle  \chi^{abc}(x_1)\chi^{abc}(x_2) \tilde{A}_s (x_3;\xi)\rangle$, we define
\begin{align}
v^{(A)}_{s,\tau}(x,y)  =  \frac{((x-y)\cdot \xi)^s}{(x-y)^{2\Delta_\phi-\tau}}, \quad v^{(C)}_{s,\tau}(x,y)  =  \frac{((x-y)\cdot \xi)^s}{(x-y)^{2\Delta_\chi-\tau}}.
\end{align}

We now define the following kernels, depicted schematically in figure \ref{kernelA}: 
\begin{eqnarray}
K_{AA}[v^{(A)}] & = & 3  \int d^d x d^d y G(x_1,x)G(x_2,y) G(x,y)F(x,y) v^{(A)}_{s,\tau}(x,y) 
 \\
K_{CA}[v^{(A)}] & = & 3 \int d^d x d^d y F(x_1,x)F(x_2,y)G(x,y)^2 v^{(A)}_{s,\tau}(x,y) 
\\
K_{AC}[v^{(C)}] & = & 3 \int d^d x d^d y G(x_1,x)G(x_2,y)G(x,y)^2 v^{(C)}_{s,\tau}(x,y) 
\end{eqnarray}
Note the factor of $3$, which appears from a careful counting of the Wick contractions. 

The integration kernel gives rise to the following matrix
\begin{equation}
\begin{pmatrix}
2 K_{AA}[v^{(A)}]/v^{(A)} & K_{AC}[v^{(C)}]/v^{(A)} \\
K_{CA}[v^{(A)}]/v^{(C)} & 0 
\end{pmatrix}
\equiv \begin{pmatrix}
2 K_1 & K_3 \\
K_2 & 0 
\end{pmatrix}\ .
\end{equation}
The condition for it to have eigenvalue 1, which determines the allowed values of $\tau$, is 
\begin{equation}
g^{(A)}(d,\tau,s)\equiv 2K_1+K_3K_2=1
\ .\end{equation}
Luckily, this condition is independent of the constant $A$, as one can see from the following expressions,
\begin{eqnarray}
K_1 & = & \frac{3 (d-6 \Delta_\phi ) \Gamma (3 \Delta_\phi ) \sin
   \left(\frac{1}{2} \pi  (d-6 \Delta_\phi )\right)
   \Gamma (d-3 \Delta_\phi ) \Gamma
   \left(\frac{d}{2}-\Delta_\phi \right)^2 \Gamma
   \left(\Delta_\phi -\frac{\tau }{2}\right) \Gamma
   \left(-\frac{d}{2}+s+\Delta_\phi +\frac{\tau
   }{2}\right)}{2 \pi  \Gamma (\Delta_\phi )^2 \Gamma
   \left(d-\Delta_\phi -\frac{\tau }{2}\right) \Gamma
   \left(\frac{1}{2} (d+2 s-2 \Delta_\phi +\tau
   )\right)} \ ,\nonumber \\
K_2 & = & \frac{3 \pi ^d 2^{4 (d-2 \Delta_\phi )} \Gamma (3 \Delta_\phi
   )^2 \Gamma \left(\frac{d}{2}-\Delta_\phi \right)^4
   \Gamma \left(d-3 \Delta_\phi -\frac{\tau }{2}\right)
   \Gamma \left(\frac{1}{2} (d+2 s-6 \Delta_\phi +\tau
   )\right)}{A^4 \lambda ^2 \Gamma (\Delta_\phi )^4
   \Gamma \left(\frac{d}{2}-3 \Delta_\phi \right)^2
   \Gamma \left(3 \Delta_\phi -\frac{\tau }{2}\right)
   \Gamma \left(-\frac{d}{2}+s+3 \Delta_\phi +\frac{\tau
   }{2}\right)}  ,\nonumber \\
K_3 & = & \frac{3 A^4 \pi ^{-d} \lambda ^2 2^{8 \Delta_\phi -4 d}
   \Gamma (\Delta_\phi )^2 \Gamma \left(\Delta_\phi
   -\frac{\tau }{2}\right) \Gamma
   \left(-\frac{d}{2}+s+\Delta_\phi +\frac{\tau
   }{2}\right)}{\Gamma \left(\frac{d}{2}-\Delta_\phi
   \right)^2 \Gamma \left(d-\Delta_\phi -\frac{\tau
   }{2}\right) \Gamma \left(\frac{1}{2} (d+2 s-2
   \Delta_\phi +\tau )\right)} \ .
\end{eqnarray}
Thus, the equation we need to solve is:
 \makeatletter 
\def\@eqnnum{{\normalsize \normalcolor (\theequation)}} 
\makeatother
{ \small
\begin{equation}
\begin{split}
& \frac{\Gamma (\Delta_\phi)^2 \Gamma
   \left(\frac{d}{2}-3 \Delta_\phi\right)^2 \Gamma
   \left(3 \Delta_\phi-\frac{d}{2}\right) \Gamma
   \left(3 \Delta_\phi-\frac{\tau }{2}\right) \Gamma
   \left(d-\Delta_\phi-\frac{\tau }{2}\right) \Gamma
   \left(-\frac{d}{2}+s+3 \Delta_\phi+\frac{\tau
   }{2}\right) \Gamma \left(\frac{1}{2} (d+2 s-2
   \Delta_\phi+\tau )\right)}{3 \Gamma (3 \Delta_\phi) \Gamma
   \left(\frac{d}{2}-\Delta_\phi\right)^2 \Gamma
   \left(\Delta_\phi-\frac{\tau }{2}\right) \Gamma
   \left(-\frac{d}{2}+s+\Delta_\phi+\frac{\tau
   }{2}\right)} \\
&=  3 \Gamma (3 \Delta_\phi) \Gamma
   \left(3 \Delta_\phi-\frac{d}{2}\right) \Gamma
   \left(d-3 \Delta_\phi-\frac{\tau }{2}\right) \Gamma
   \left(\frac{1}{2} (d+2 s-6 \Delta_\phi+\tau
   )\right)- \\
    &-2 \Gamma \left(\frac{d}{2}-3 \Delta_\phi
   \right) \Gamma (d-3 \Delta_\phi) \Gamma \left(3
   \Delta_\phi-\frac{\tau }{2}\right) \Gamma
   \left(-\frac{d}{2}+s+3 \Delta_\phi+\frac{\tau
   }{2}\right)\ .
\end{split}
\end{equation}}
One can check that the stress-tensor, which has $s=2$ and $\tau=d-2$, appears in this spectrum for any $d$. 

\begin{figure}[h!]
\begin{center}
\includegraphics[width=10cm]{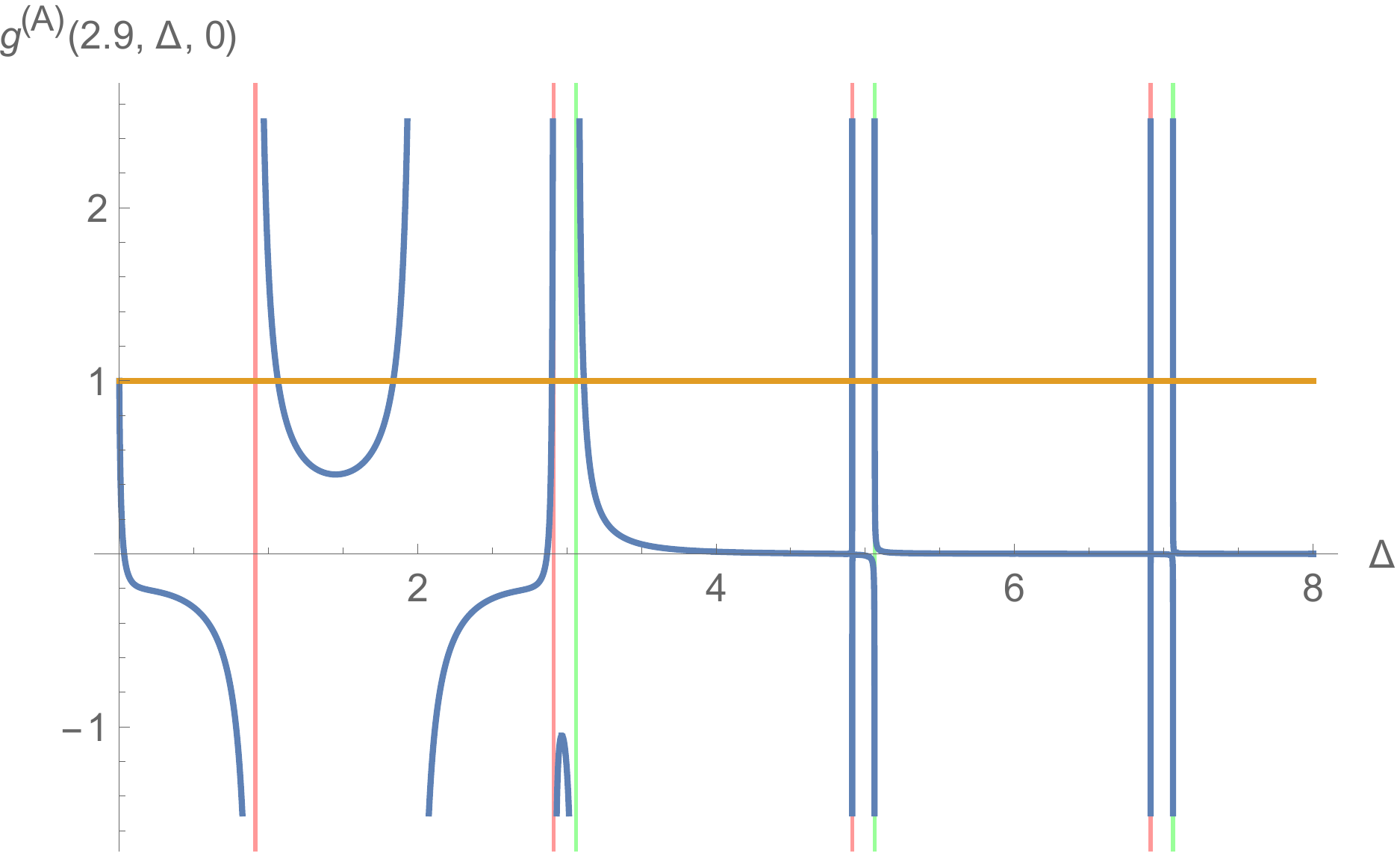}
\caption{The spectrum of type A/C scalar bilinears in $d=2.9$. The green lines correspond to the $2\Delta_\chi+2n$ asymptotics and the red ones to $2\Delta_\phi+2n$ asymptotics.
We see that the solutions are real, and approach the expected values as $n \rightarrow \infty$. \label{typeA2_9} }
\end{center}
\end{figure}

\begin{figure}[h!]
\begin{center}
\includegraphics[width=10cm]{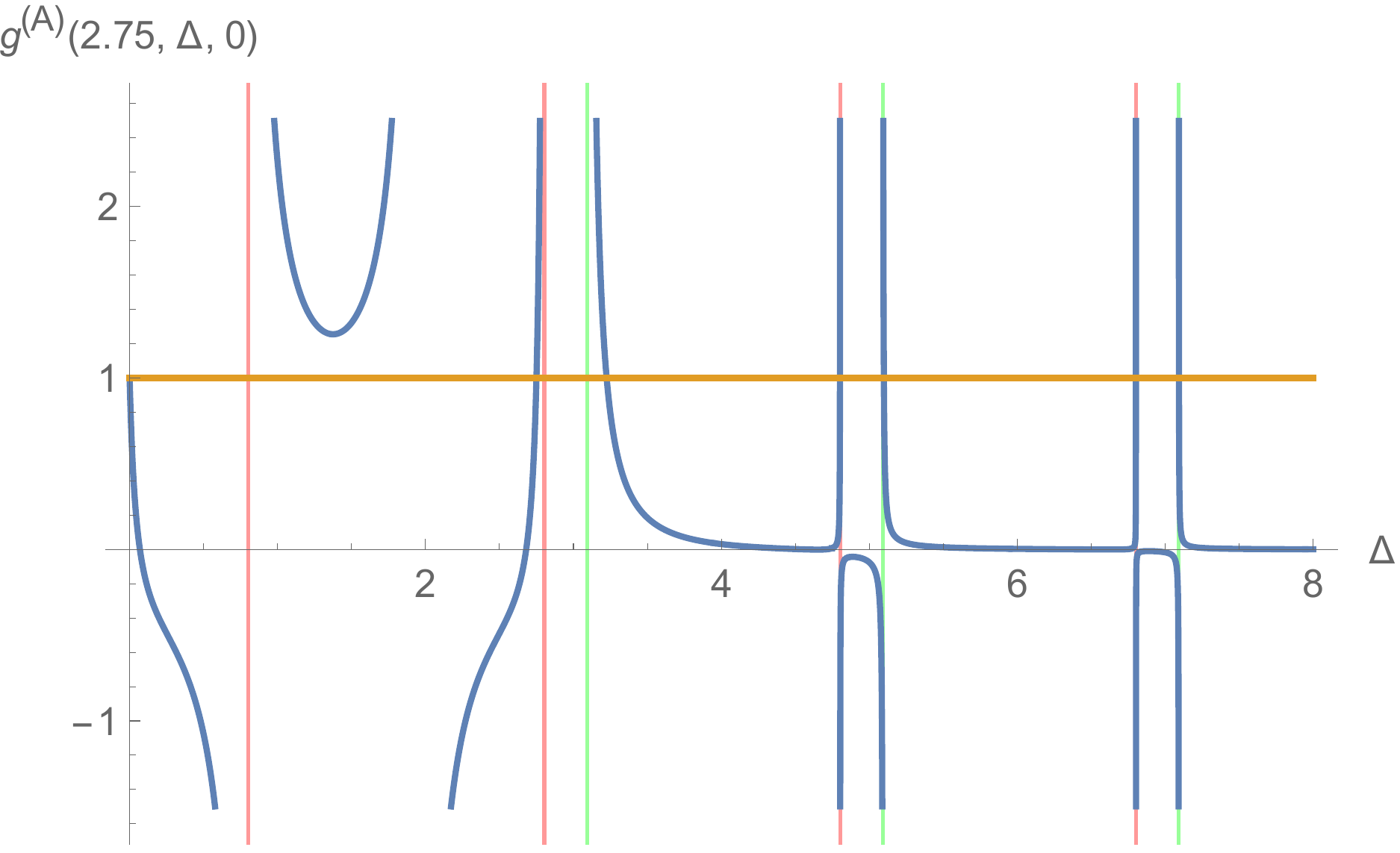}
\caption{The spectrum of type A/C scalar bilinears in $d=2.75$. The green lines correspond to the $2\Delta_\chi+2n$ asymptotics and the red ones to $2\Delta_\phi+2n$ asymptotics. 
We see that two real solutions are no longer present; they are now complex. \label{typeA2_75} }
\end{center}
\end{figure}

The Schwinger-Dyson equations have a symmetry under $\Delta \rightarrow d-\Delta$.
In a given CFT, only one of this pair of solutions corresponds to a primary operator dimension, while the other one is its ``shadow."
The $s=0$ spectrum contains complex modes for $1.6799<d<2.8056$. 
In $d=2.9$ 
the graphical solution for the scaling dimensions in the type A/C sector is shown in figure \ref{typeA2_9}.
The lowest few are 
\begin{equation}
\Delta=1.064,~1.836,~2.9,~3.114,~4.912,~5.063,~6.913, ~7.063, \ldots
\end{equation}
The eigenvalue at $\Delta=2.9$ is exact, and in general $\Delta=d$ is an eigenvalue for any $d$. The solution $1.836$ corresponds 
to the shadow of  $1.064$.
As $d$ is further lowered, the part of the graph between $1$ and $2$ moves up so that the two solutions become closer. 
In $d=d_{\rm crit}$, where $d_{\rm crit} \approx 2.8056$, the two solutions merge into a single one at $d/2$
(for discussions of mergers of fixed points, see \cite{Kaplan:2009kr,Giombi:2015haa,Gorbenko:2018ncu}).
For $d<  d_{\rm crit}$, the solutions become complex $\frac{d}{2} \pm i \alpha(d)$ and the prismatic model becomes unstable.
 The plot for $d=2.75$ 
is shown in figure \ref{typeA2_75}.

\begin{figure}[h!]
\begin{center}
\includegraphics[width=10cm]{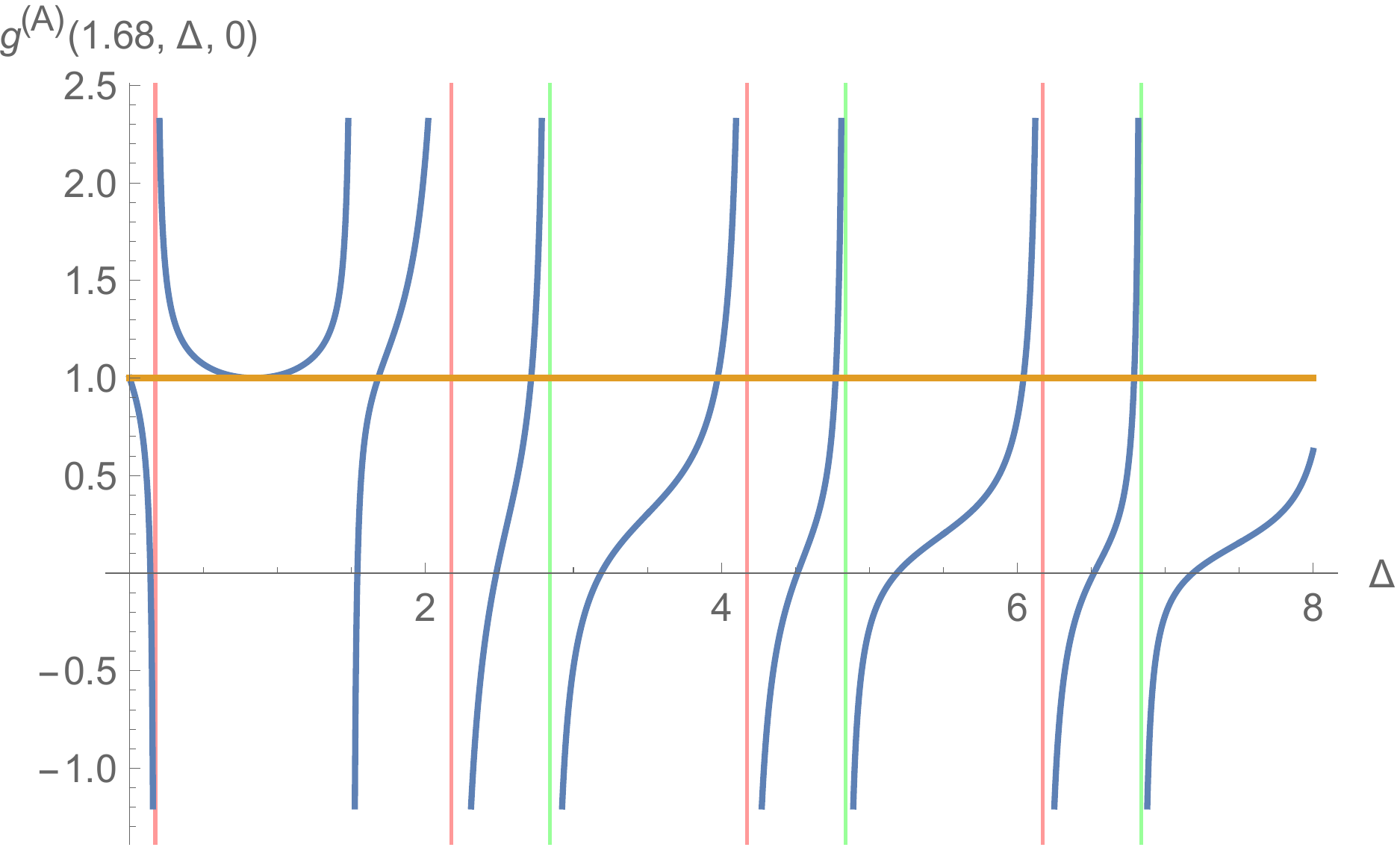}
\caption{The spectrum of type A/C scalar bilinears in $d=1.68$. The green vertical lines correspond to the $2\Delta_\chi+2n$ asymptotics; 
the red ones to the $2\Delta_\phi+2n$ asymptotics. \label{typeA1_6} }
\end{center}
\end{figure}

For $d\leq 1.68$, the spectrum of bilinears is again real. 
The plot for $d=1.68$, where $\Delta_\phi\approx 0.0867$, 
is shown in figure \ref{typeA1_6}. At this critical value of $d$ there are two solutions at $d/2$; one is the shadow of the other.

\section{Large $N$ results in $3-\epsilon$ dimensions}
\label{threeeps}

Let us solve the Schwinger-Dyson equations in $d=3-\epsilon$. The results will be compared with renormalized perturbation theory in the following section.
The scaling dimension of $\phi^{abc}$ is found to be
\begin{equation}
\Delta_\phi=\frac{1}{2}-\frac{\epsilon }{2}+\epsilon ^2-\frac{20
   \epsilon ^3}{3}+\left(\frac{472}{9}+\frac{\pi
   ^2}{3}\right) \epsilon ^4+\left(7 \zeta
   (3)-\frac{12692}{27}-\frac{56 \pi ^2}{9}\right)
   \epsilon ^5+\mathcal{O}\left(\epsilon ^6\right)\ .
\label{SDdim}
\end{equation}
This is within the allowed range (\ref{allowedrange}) and is close to its upper boundary.
The scaling dimension of $\chi^{abc}$ is
\begin{equation}
\Delta_\chi=d- 3 \Delta_\phi= \frac{3}{2}+ \frac{\epsilon }{2}- 3\epsilon ^2 + 20
   \epsilon ^3- \left(\frac{472}{3}+\pi^2 \right) \epsilon ^4 - 3\left(7 \zeta
   (3)-\frac{12692}{27}-\frac{56 \pi ^2}{9}\right)
   \epsilon ^5+\mathcal{O}\left(\epsilon ^6\right)\ .
\label{SDchidim}
\end{equation}

Let us consider the $s=0$ type A/C bilinears. For the first eigenvalue we find, 
\begin{equation}
\Delta_{\phi^2} = 1-\epsilon +32 \epsilon ^2-\frac{976 \epsilon
   ^3}{3}+\left(\frac{30320}{9}+\frac{32 \pi
   ^2}{3}\right) \epsilon ^4+
\mathcal{O}\left(\epsilon ^5\right)\ .
\label{SDphi2dim}
\end{equation}
It corresponds to the scaling dimension of operator
$\phi^{abc} \phi^{abc}$, as we will show in the next section.
The next eigenvalue is the shadow dimension $d-  \Delta_{\phi^2}$.

The next solution of the S-D equation is $\Delta=d=3-\epsilon$ for all $d$.
While this seems to correspond to an exactly marginal operator, we believe that the corresponding operator is redundant: it is a linear combination of
 $\phi^{abc}\partial^2 \phi^{abc}$ and $\chi^{abc} \chi^{abc}$. Similar redundant operators with $h=1$ showed up in the Schwinger-Dyson analysis of multi-flavor
models \cite{Gross:2016kjj, Bulycheva:2017ilt}. They decouple in correlation functions \cite{Gross:2016kjj} 
and were shown to vanish by the equations of motion \cite{Bulycheva:2017ilt}. 
The next eigenvalue is 
\begin{equation}
\Delta_{\rm prism}=3+\epsilon +6\epsilon^2-84 \epsilon^3+\left(\frac{1532}{3}+10 \pi ^2\right) \epsilon^4+\left(18 \zeta (3)-\frac{6392}{3}-\frac{452 \pi^2}{3}\right) \epsilon^5+
\mathcal{O}\left(\epsilon^6\right)\ .
\label{SDprismdim}
\end{equation}
It should correspond to the sextic prism operator (\ref{prism}), which is related by the equations of motion to a linear combination of 
$\phi^{abc}\partial^2 \phi^{abc}$ and $\chi^{abc} \chi^{abc}$.

The subsequent eigenvalues may be separated into two sets. One of them has the form, for integer $n\geq 0$,
\begin{align}
\Delta_n^- =&  5+2n-\epsilon +2 \epsilon ^2-\frac{40 \epsilon
   ^3}{3}+\notag \\ 
&+\frac{\left(2 \left(472+3 \pi ^2\right) n
   (2 n+7) (n (2 n+7)+11)+180 \pi ^2+28212\right)
   \epsilon ^4}{9 (n+1) (n+2) (2 n+3) (2
   n+5)}+\mathcal{O}\left(\epsilon ^5\right).
\end{align}
For large $n$ this approaches $4+ 2n + 2\Delta_\phi$, as expected for an operator of the form  $\phi^{abc}(\partial^2)^{2+n} \phi^{abc}$.
The other set of eigenvalues has the form, for integer $n\geq 0$,
\begin{equation}
\Delta_n^+= 5+2n +\epsilon - 6 \epsilon ^2 +4 \left(\frac{9}{n+2}-\frac{18}{2 n+3}-\frac{6}{2
   n+5}+\frac{3}{n+1}+10\right) \epsilon
   ^3+\mathcal{O}\left(\epsilon
   ^4\right)\ .
\end{equation}
For large $n$ this approaches $2+ 2n + 2\Delta_\chi$, as expected for an operator of the form  $\chi^{abc}(\partial^2)^{1+n} \chi^{abc}$.
These simple asymptotic forms suggest that for large $n$ the mixing between operators $\phi^{abc}(\partial^2)^{2+n} \phi^{abc}$
and $\chi^{abc}(\partial^2)^{1+n} \chi^{abc}$ approaches zero.

We can also use (\ref{typeBscalars}) to derive the $3-\epsilon$ expansions of the dimensions of type B operators, 
\begin{equation}
O_{B,n}=
\chi^{abc} (\partial_\mu \partial^\mu)^n \phi^{abc}+ \ldots \ ,
\end{equation}
where the additional terms are there to make them conformal primaries.
For $n=0$ we find
\begin{equation}
\Delta_{B,0}= 2 + 6\epsilon -68 \epsilon ^2+ \frac{2848 + 24\pi^2}{3} \epsilon^3
  +\mathcal{O}\left(\epsilon^4\right)\ .
\label{SDtetra}
\end{equation}
This scaling dimension corresponds to the operator $\phi^{abc}\chi^{abc}$, which in the original $\phi$ language is the tetrahedron operator
$O_{\rm tetra}$.
For the higher operators we get
 \begin{align}
& \Delta_{B,1}=4+ 4 \epsilon^3
- 44\epsilon^4
  +\mathcal{O}\left(\epsilon^5\right)\ ,\\
& \Delta_{B,2}=6-\frac{7}{5} \epsilon^2+ \frac{331}{30} \epsilon^3
-\left(\frac{199547}{2250}+\frac{7 \pi ^2}{15}\right) \epsilon^4
  +\mathcal{O}\left(\epsilon^5\right)\ , \\ 
& \Delta_{B,3}=8 -\frac{12}{7}\epsilon^2+  \frac{9139}{735} \epsilon^3
-\left(\frac{7581556}{77175}+\frac{4 \pi ^2}{7}\right) \epsilon^4
  +\mathcal{O}\left(\epsilon^5\right)\ ,
\ {\rm etc.}
\label{SDtetrahigher}
\end{align}
Using the equations of motion, we can write $O_{B,1}$, up to a total derivative, as 
a sum of the three $8$-particle operators shown in the leftmost column of figure 9 in \cite{Bulycheva:2017ilt}. 
In general, for $n>0$, 
\begin{align}
\Delta_{B,n}=2n+2 -2 \left (1- \frac{3}{n(2n+1)} \right )\epsilon^2 + \mathcal{O}\left(\epsilon^3\right)\ ,
\end{align}
which agrees for large $n$ with the expected asymptotic behavior 
\begin{align}
\Delta_{B,n}\rightarrow 2n+\Delta_\phi+ \Delta_\chi= 2n+ 2 -2 \epsilon^2 + \mathcal{O}\left(\epsilon^3\right)\ .
\end{align}

\subsection{Higher Spin Spectrum}

Let us also present the $\epsilon$ expansions for 
the higher spin bilinear operators which are mixtures of type $A$ and $C$. The lowest eigenvalue of twist $\tau=\Delta-s$ for spin $s$ is
\begin{align}
\tau_0  =&1 -\epsilon +\frac{8 \left(s^2-4\right) \epsilon ^2}{4
   s^2-1}\notag \\
&+\frac{4 \epsilon ^3 \left(27 \left(1-4
   s^2\right) H_{s-\frac{1}{2}}-2 s \left(80 s^3+s
   (54 \log (4)-508)+45\right)-244+27 \log
   (4)\right)}{3 \left(1-4
   s^2\right)^2}+\mathcal{O}\left(\epsilon ^4\right)
\end{align}
where $H_{n}$ is the harmonic number and the last two terms (as well as all higher-order terms) vanish when $s=2$ as expected. In the large $s$ limit, this becomes:
\begin{equation}
\begin{aligned}
\tau_0&\rightarrow 1-\epsilon +\epsilon ^2 \left(2-\frac{15}{2
   s^2}+ \mathcal{O}(s^{-3})
\right)+
   \epsilon ^3 \left(-\frac{40}{3}+\frac{-9 \log (4 s)-9
   \gamma
   +78}{s^2}+ \mathcal{O}(s^{-3})
\right)
   +\mathcal{O}\left(\epsilon ^4\right)\,.
\end{aligned}
\end{equation}
Comparing with (\ref{deltaphi}), we see that  
\begin{equation}
\tau_0 = 2\Delta_{\phi}+O(\frac{1}{s^2})\,
\end{equation} 
This is the expected large spin limit \cite{Parisi:1973xn, Callan:1973pu, Fitzpatrick:2012yx, Komargodski:2012ek} for an operator bilinear in $\phi$, indicating that for large spin the mixing with $\chi$ bilinears is suppressed. 

The next two twists are
\begin{align}
\tau_1=& 3-\epsilon +\frac{8 s (s+2) \epsilon ^2}{4 s
   (s+2)+3} 
\notag \\
& +\frac{4 \epsilon ^3}{3 (4 s
   (s+2)+3)^2} \bigg(-4 (40 s
   (s+4)+157) s^2+6 (s+27)-27 \gamma  (4 s
   (s+2)+3)
\notag \\
&-27 (4 s (s+2)+3) \log (4)-27 (4 s
   (s+2)+3) \psi (s+\frac{3}{2})\bigg)
+\mathcal{O}\left(\epsilon ^4\right)\ ,
\end{align} 
and 
\begin{align}
\tau_2 =&3+\epsilon +\left(\frac{36}{4 s (s+2)+3}-6\right)
   \epsilon ^2  \notag \\
&+\frac{4 \epsilon ^3}{(4 s (s+2)+3)^2} \bigg(4 s (2 s (20 s (s+4)+56+9 \log
   4)-105+36 \log 4)\notag \\
&+18 \gamma 
   (4 s (s+2)+3)+18 (4 s (s+2)+3) \psi (s+\frac{3}{2})-297+54 \log
   4\bigg) +\mathcal{O}\left(\epsilon^4\right)\,,
\end{align}
where $\psi(x)$ is the digamma function. 
In the large $s$ limit, these take the form,
\begin{equation}
\begin{aligned}
\tau_1&\rightarrow 3-\epsilon +\epsilon ^2 \left(2-\frac{3}{2
   s^2}+ O(s^{-3})
\right)+
   \epsilon ^3 \left(-\frac{40}{3}-\frac{3 (3 \log
   (s)+\log (64)+3 \gamma
   -7)}{s^2}+ O(s^{-3})
\right)+O\left(\epsilon ^4\right)\\
&=2\Delta_{\phi}+2+O(\frac{1}{s^2})\,,
\end{aligned}
\end{equation}
and
\begin{equation}
\begin{aligned}
\tau_2&\rightarrow 3+\epsilon +\epsilon ^2
   \left(-6+\frac{9}{s^2}+ O(s^{-3})
\right)+\epsilon ^3 \left(40+\frac{18
   (\log (s)+\log (4)+\gamma
   -6)}{s^2}+ O(s^{-3})
   \right)+O\left(\epsilon ^4\right)\\
&=2\Delta_{\chi}+O(\frac{1}{s^2})\,,
\end{aligned}
\end{equation}
In general, for large 
spin we find the two towers of twists labelled by an integer $n$
\begin{equation}
\begin{aligned}
&\tau_{n}^A = 2 n+1-\epsilon +2 \epsilon ^2-\frac{40 \epsilon
   ^3}{3}+\mathcal{O}(\epsilon ^4) = 2\Delta_{\phi}+2n+\ldots \\
&\tau_{n}^C = 2 n+3+\epsilon -6 \epsilon ^2+\mathcal{O}(\epsilon
   ^3) = 2\Delta_{\chi}+2n+\ldots 
\end{aligned}
\end{equation}
again in agreement with the expected asymptotics and suppression of mixing at large spin. 


We can similarly derive explicit results for spinning operators in the type B sector using (\ref{hsBeq}). For the lowest two twists, we find
\begin{equation}
\begin{aligned}
\tau_0 &=2+\frac{6 \epsilon }{2 s+1}+\frac{2 \epsilon ^2 \left(3 (2 s+1)^2 \left(H_{s-\frac{1}{2}}+\log (4)\right)-8 s^3-84 s^2-72 s-34\right)}{(2 s+1)^3}+O(\epsilon^3) \\
&=\left(2-2\epsilon^2+O(\epsilon^3)\right)+O(\frac{1}{s})\,,\\
\tau_1 &= 4-\frac{4 s \epsilon ^2}{2 s+3} + \frac{2 \epsilon ^3 \left(9 (2 s+3) H_{s+\frac{1}{2}}+80 s^2+12 s (8+\log (8))+54 \log (2)\right)}{3 (2 s+3)^2}+O(\epsilon^4)\\
&=\left(4-2 \epsilon ^2+\frac{40 \epsilon ^3}{3}+O(\epsilon^4)\right)+O(\frac{1}{s})\,,
\end{aligned}
\end{equation}
and higher twists may be analyzed similarly. 
One can see that these results are also in agreement with the expected large spin limit $\tau_n \rightarrow \Delta_{\phi}+\Delta_{\chi}+2n$ for fixed $n$.

\section{Renormalized perturbation theory}
\label{threeminuseps}

In this section we use the renormalized perturbation theory to carry out the $3-\epsilon$ expansion for finite $N$.
We will find a fixed point with real couplings, whose large $N$ limit reproduces the results found using the $3-\epsilon$ expansion of the 
Schwinger-Dyson solution in the previous section.
This is an excellent check of the Schwinger-Dyson approach to the prismatic theory.

To carry out the beta function calculation at finite $N$ we need to include all the $O(N)^3$ invariant sextic terms in the action 
(as usual in such calculations, we ignore the quartic and quadratic operators which are relevant in $d=3$).
The 11 such single-sum terms are shown diagrammatically in figure 5 of \cite{Bulycheva:2017ilt}.
We will impose the additional constraint that the action is invariant under the permutation group $S_3$ which acts on the three $O(N)$ symmetry groups.
This leaves us with 8 operators: 5 single-sum, 2 double-sum and 1 triple-sum. 
They are written down explicitly in
(\ref{allinter}) and 
shown schematically in figure \ref{8inter}.
The first one and the most essential one for achieving the solvable large $N$ limit is the ``prism" term (\ref{prism});
it is positive definite and symmetric under the interchanges of the three $O(N)$ groups.

\begin{figure}
	\centering
	\includegraphics[scale=0.8]{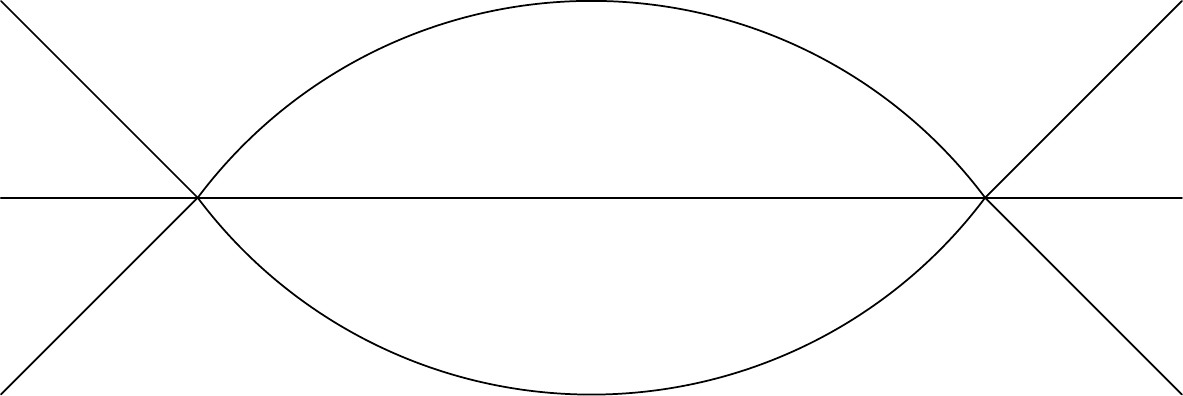}
	\caption{\label{fig:2loopdig} The two-loop contribution to the beta-function.}
\end{figure}

Our action is a special case of a general multi-field $\phi^6$ tensor theory:
\begin{align}
S = \int d^{d}x \left(\frac{1}{2}\partial_{\mu}\phi^{abc}\partial^{\mu}\phi^{abc}+\frac{1}{6!}g_{\kappa_{1}\kappa_{2}\kappa_{3}\kappa_{4}\kappa_{5}\kappa_{6}}\phi^{\kappa_{1}}\phi^{\kappa_{2}}\phi^{\kappa_{3}}\phi^{\kappa_{4}}\phi^{\kappa_{5}}\phi^{\kappa_{6}}\right)\, .
\end{align}
The beta-functions and anomalous dimensions for such a general sextic coupling were calculated in \cite{Gracey:2015fia,Osborn:2017ucf};
see also \cite{Pisarski:1982vz,Hager:2002uq} for earlier results on the $O(n)$ invariant sextic theory.
The diagram topology contributing to the leading two-loop beta function 
is shown in figure \ref{fig:2loopdig}. 

In our case each index $\kappa_{1}, \kappa_{2}\dots, \kappa_{6}$ has three sub indices $\kappa_{i}=(a_{i}b_{i}c_{i})$. The coupling $g_{\kappa_{1}\kappa_{2}\kappa_{3}\kappa_{4}\kappa_{5}\kappa_{6}}$ contains $8$ different types of interactions
\begin{align}
g_{\kappa_{1}\kappa_{2}\kappa_{3}\kappa_{4}\kappa_{5}\kappa_{6}} = g_{1}T^{(1)}_{\kappa_{1}\kappa_{2}\kappa_{3}\kappa_{4}\kappa_{5}\kappa_{6}}+g_{2}T^{(2)}_{\kappa_{1}\kappa_{2}\kappa_{3}\kappa_{4}\kappa_{5}\kappa_{6}}+\dots +g_{8}T^{(8)}_{\kappa_{1}\kappa_{2}\kappa_{3}\kappa_{4}\kappa_{5}\kappa_{6}}\,,
\end{align}
which can be graphically represented as in figure \ref{8inter}.
Each tensor structure $T^{(k)}_{\kappa_{1}\kappa_{2}\kappa_{3}\kappa_{4}\kappa_{5}\kappa_{6}}$ consists of a sum of  product of $\delta$ functions, which are symmetrized over the colors $(abc)$ and over the indices $\kappa_{1},\dots, \kappa_{6}$.

The two-loop beta functions and anomalous dimensions for general $N$ are given in the Appendix. 
Let us use the large $N$ scaling 
\begin{align}
&g_{1}= 180 \cdot (8\pi)^{2}\epsilon\frac{\tilde{g}_{1}}{N^{3}}\ , \qquad g_{2,4,6,7}= 180\cdot (8\pi)^{2}\epsilon\frac{\tilde{g}_{2,4,6,7}}{N^{5}}\ , 
\notag \\
& g_{3,5}= 
180\cdot (8\pi)^{2}\epsilon\frac{\tilde{g}_{3,5}}{N^{4}}\ , \qquad g_{8}= 180\cdot (8\pi)^{2}\epsilon\frac{\tilde{g}_{8}}{N^{7}}\ , 
\label{scalings}
\end{align}
which is chosen in such a way that all beta functions retain non-vanishing quadratic terms 
in the large $N$ limit: 
\begin{align}
&\tilde{\beta}_{1}= -2 \tilde{g}_1+2 \tilde{g}_1^2\ , \quad\tilde{\beta}_{2}= -2 \tilde{g}_2+ 4 \tilde{g}_1 \left(3 \tilde{g}_1+2 \tilde{g}_5\right)\ , \quad \tilde{\beta}_{3}= -2  \tilde{g}_3+12 \tilde{g}_1^2\ , \notag\\
&\tilde{\beta}_{4}=-2  \tilde{g}_4+\frac{2}{3} \left(2 \left(3 \tilde{g}_1+\tilde{g}_3\right)^2+\tilde{g}_5^2+12 \tilde{g}_1 \tilde{g}_5\right)\ , \quad 
\tilde{\beta}_{5}=-2  \tilde{g}_5+4\tilde{g}_1 \left(6 \tilde{g}_1+\tilde{g}_5\right)\ , \notag\\
&\tilde{\beta}_{6}=-2   \tilde{g}_6+4\tilde{g}_1 \left(3 \tilde{g}_1+\tilde{g}_5+2 \tilde{g}_6\right), \quad \tilde{\beta}_{7}= -2  \tilde{g}_7+6\tilde{g}_1^2\ , \notag\\
&\tilde{\beta}_{8}=-2  \tilde{g}_8+\frac{4}{3} \left(\tilde{g}_3^2+4 \tilde{g}_7 \tilde{g}_3+\tilde{g}_5^2+6 \tilde{g}_6^2+2 \tilde{g}_7^2+6 \tilde{g}_5 \tilde{g}_6+3 \tilde{g}_1 \left(\tilde{g}_5+6 \tilde{g}_6\right)\right)\ .
\end{align}
The unique non-trivial fixed point of these scaled beta functions is at
\begin{align}
& \tilde{g}_1^* = 1 , \quad
\tilde{g}_2^* = -42 , \quad
\tilde{g}_3^* = 6 , \quad
\tilde{g}_4^* = 54  , \notag \\
& \tilde{g}_5^* = -12 , \quad
\tilde{g}_6^* = 6 , \quad
\tilde{g}_7^* = 3 , \quad
\tilde{g}_8^* = 84 .
\label{uniquefixed}
\end{align}
For this fixed point, the eigenvalues of the matrix $\frac{\partial \tilde{\beta}_i}{\partial \tilde{g}_j}$ are 
\begin{align}
\lambda_{i}=6,~2,~2,~-2,~-2,~-2,~-2,~-2\ .
\end{align} 
That there are unstable directions at the ``prismatic" fixed point also follows from the solution of the Schwinger-Dyson equations.\footnote{
At finite $N$, using the beta functions given in the Appendix, we are able to find and study additional fixed points numerically. The analysis of behavior of the beta-functions shows that 
they are all saddle points and, therefore, neither stable in the IR nor in the UV.} 
Using (\ref{SDphi2dim}) we see that
the large $N$ dimension of the triple-trace operator $(\phi^{abc}\phi^{abc})^3$ is 
$3(1-\epsilon)+ \mathcal{O}(\epsilon^2)$, which means that it is relevant in $d=3-\epsilon$ and is one of the operators corresponding to eigenvalue $-2$.
On the other hand, the prism operator is irrelevant and corresponds to eigenvalue $2$. Another irrelevant operator is $O_{\rm tetra} \phi^{abc}\phi^{abc}$;
from (\ref{SDtetra}) it follows that
its large $N$ dimension is $3+5\epsilon + \mathcal{O}(\epsilon^2)$, so it corresponds to eigenvalue $6$. 

We have also calculated the $1/N$ corrections to the fixed point (\ref{uniquefixed}):
\begin{align}
& \tilde{g}_1^* = 1- \frac{6}{N} + \frac{18}{N^2} +\ldots , \notag \\
&\tilde{g}_2^* = -42 + \frac{384}{N} + \frac{8592}{N^2} +\ldots  , \notag \\
&\tilde{g}_3^* = 6+  \frac{1848}{N^2} +\ldots  , \notag \\
&\tilde{g}_4^* = 54 - \frac{132}{N} + \frac{16392}{N^2} +\ldots  , \notag \\
& \tilde{g}_5^* = -12+ \frac{30}{N} + \frac{2340}{N^2} +\ldots  , \notag \\
&\tilde{g}_6^* = 6+ \frac{36}{N} -\frac{1320}{N^2} +\ldots  , \notag \\
&\tilde{g}_7^* = 3 + \frac{174}{N} + \frac{7080}{N^2} +\ldots , \notag \\
&\tilde{g}_8^* = 84 + \frac{6732}{N} + \frac{309204}{N^2} +\ldots 
\label{uniqcorr}
\end{align}
For the scaling dimension of $\phi$, we find from 
(\ref{gammaphi}):
\begin{align}
\Delta_\phi= \frac{d-2}{2}+ \gamma_\phi = \frac{1}{2}-\frac{\epsilon }{2}+ \epsilon^2 \left ( 1- \frac{12}{N} + \frac{75}{N^2}+ \ldots \right )+ \mathcal{O}(\epsilon^3)\, .
\label{deltaphi}
\end{align}
In the large $N$ limit, (\ref{deltaphi}) is
in agreement with the solution of the S-D equation (\ref{SDdim}). 
For the scaling dimension of $\phi^{abc} \phi^{abc}$, we find 
\begin{align}
\Delta_{\phi^2} = d-2+ \gamma_{\phi^2} =1-\epsilon + 32 \epsilon^2 \left ( 1- \frac{12}{N} + \frac{75}{N^2}+ \ldots \right ) + \mathcal{O}(\epsilon^3)\,.
\label{deltaphisquared}
\end{align}
In the large $N$ limit
this is in agreement with (\ref{SDphi2dim}).
In general, calculating the $1/N$ corrections in tensor models seems to be quite difficult \cite{Gurau:2016lzk}, but it is nice to
see that in the prismatic QFT the $3-\epsilon$ expansion provides us with explicit results for the $1/N$ corrections to scaling dimensions of various operators.

The scaling dimension of the marginal prism operator is
\begin{equation}
\Delta_{\rm prism} = d+ {d\tilde \beta_1\over d\tilde g_1}= 3-\epsilon - 2\epsilon+ 4\epsilon \tilde g_1^*+ \ldots= 3+ \epsilon + \mathcal{O}(\epsilon^2)\ ,
\end{equation}
which is in agreement with (\ref{SDprismdim}).

We have also performed two-loop calculations of the scaling dimensions of the tetrahedron and pillow operators; see the appendix for the
anomalous dimension matrix.
In the large $N$ limit, we find
\begin{align}
& \Delta_{\rm tetra} = 2(d-2) + \gamma_{\rm tetra}= 2+ 6\epsilon + \mathcal{O}(\epsilon^2)\ , \notag \\
& \Delta_{\rm pillow} = 2(d-2) + \gamma_{\rm pillow}= 2- 2\epsilon + \mathcal{O}(\epsilon^2)\ ,
\label{tetrapillow}
\end{align}
which is in agreement with the S-D result (\ref{SDtetra}).
Thus, we see that the large $N$ $3-\epsilon$ expansions from the Schwinger-Dyson approach have passed a number of 2-loop consistency checks.

We have also solved the equations for the fixed points of two-loop beta functions numerically for finite $N$. The results 
for the prismatic fixed point are shown in table
\ref{NumSol}.
These results are in good agreement with the analytic $1/N$ expansions (\ref{uniqcorr}) for $N\geq 200$.
\begin{table}[!h!]
	\begin{center}
		\begin{tabular}{|c|c|c|c|c|c|c|c|c|c|}
			\hline
			$N$  & $\tilde{g}_1^{*}$ & $\tilde{g}_2^*$ & $\tilde{g}_3^*$ & $\tilde{g}_4^*$ & $\tilde{g}_5^*$& $\tilde{g}_6^*$ & $\tilde{g}_7^*$& $\tilde{g}_8^*$ & $\gamma_\phi/\epsilon^2$ \\
			\hline
		54 & 0.89 & -33.06 & 7.87 & 83.69 & -11.13 & 6.86 & 27.37 &
		2047.16 & 0.80\\
				\hline
			100 & 0.94 & -37.56 & 6.23 & 55.35 & -11.53 & 6.28 & 5.98
				& 212.08 & 0.89\\
				\hline
			200 & 0.97 & -39.90 & 6.05 & 53.8 & -11.80 & 6.15 & 4.09 &
			127.90 & 0.94\\
			\hline
			400 & 0.99 & -40.99 & 6.01 & 53.78 & -11.91 & 6.08 & 3.48 &
				103.03 & 0.97\\
			\hline
			2000 & 1.00 & -41.81 & 6.00 & 53.94 & -11.98 & 6.02 & 3.09 &
				87.45 & 0.99\\
			\hline
			5000 & 1.00 & -41.92 & 6.00 & 53.97 & -11.99 & 6.01& 3.04 &
				85.36 & 0.998\\
			\hline
			10000 & 1.00 & -41.96 & 6.00 & 53.99 & -12.00 & 6.00 & 3.02 & 
				84.68 & 0.999\\
			\hline
			100000 & 1.00 & -42.00 & 6.00 & 54.00 & -12.00 & 6.00 & 3.00 & 84.07 & 1.00 \\
			\hline
		\end{tabular}
	\caption{\label{NumSol} The numerical solutions for the coupling constants defined in \eqref{scalings}}
	\end{center}
\end{table}
At $N=N_{\rm crit}$, where $N_{\rm crit}\approx 53.65$, the prismatic fixed point in $3-\epsilon$ dimensions merges with another fixed point;\footnote{
This is similar, for example, to the situation in the $O(N)$ invariant cubic theory in $6-\epsilon$ 
dimensions \cite{Fei:2014yja,Fei:2014xta},
where  $N_{\rm crit}\approx 1038.266$. For general discussions of mergers of fixed points, see \cite{Kaplan:2009kr,Gorbenko:2018ncu}.}
 they are located at
\begin{align}
&\tilde{g}_1^* = 0.89  , \quad
\tilde{g}_2^* = -32.90, \quad
\tilde{g}_3^* = 8.24 , \quad
\tilde{g}_4^* = 92.01  , \notag \\
&\tilde{g}_5^* = -11.15, \quad
\tilde{g}_6^* = 7.00 , \quad
\tilde{g}_7^* = 35.33 , \quad
\tilde{g}_8^* = 3155.29\ .
\end{align}
For  $N<N_{\rm crit}$ both of them become complex.
For example, for $N=53.6$ the two complex fixed points are at
\begin{gather}
\tilde{g}_1^* = 0.89 - 0.0002 i , \quad
\tilde{g}_2^* = -32.89+0.04 i , \quad
\tilde{g}_3^* = 8.24 +0.15 i, \quad
\tilde{g}_4^* = 91.98 +3.51 i  , \quad\notag\\
\tilde{g}_5^* = -11.15-0.01 i , \quad
\tilde{g}_6^* = 7.00 +0.06 i, \quad
\tilde{g}_7^* = 35.19 +3.61 i , \quad
\tilde{g}_8^* = 3107.77 +554.01 i  
\end{gather}
and at the complex conjugate values.

\section{Bosonic Quantum Mechanics}
\label{BosonicQM}

The action (\ref{prism}) for $d=1$ describes the quantum mechanics of a particle moving in $N^3$ dimensions with a non-negative sextic potential 
which vanishes at the origin.\footnote{
A very similar $d=1$ model with a stable sextic potential was studied in \cite{Azeyanagi:2017drg,Azeyanagi:2017mre} 
using the formulation \cite{Ferrari:2017ryl} where a rank-3 tensor is viewed as $D$ matrices. It was argued
\cite{Azeyanagi:2017drg,Azeyanagi:2017mre} that the sextic bosonic model does not have a good IR limit. 
We, however, don't find an obvious problem with the prismatic $d=1$ model because the complex scaling dimensions are absent for the bilinear operators.
We note that the negative 
scaling dimension (\ref{QMsol}), which we find for $\phi$, is quite far from 
the $1/6$ mentioned in \cite{Azeyanagi:2017drg,Azeyanagi:2017mre}.}
Such a problem should exhibit a discrete spectrum with positive energy levels, and it is conceivable that
in the large $N$ limit the gaps become exponentially small, leading to a nearly conformal behavior. For moderate values of $N$, this quantum
mechanics problem may even be accessible to numerical studies.  

\begin{figure}[h!]
\begin{center}
\includegraphics[width=10cm]{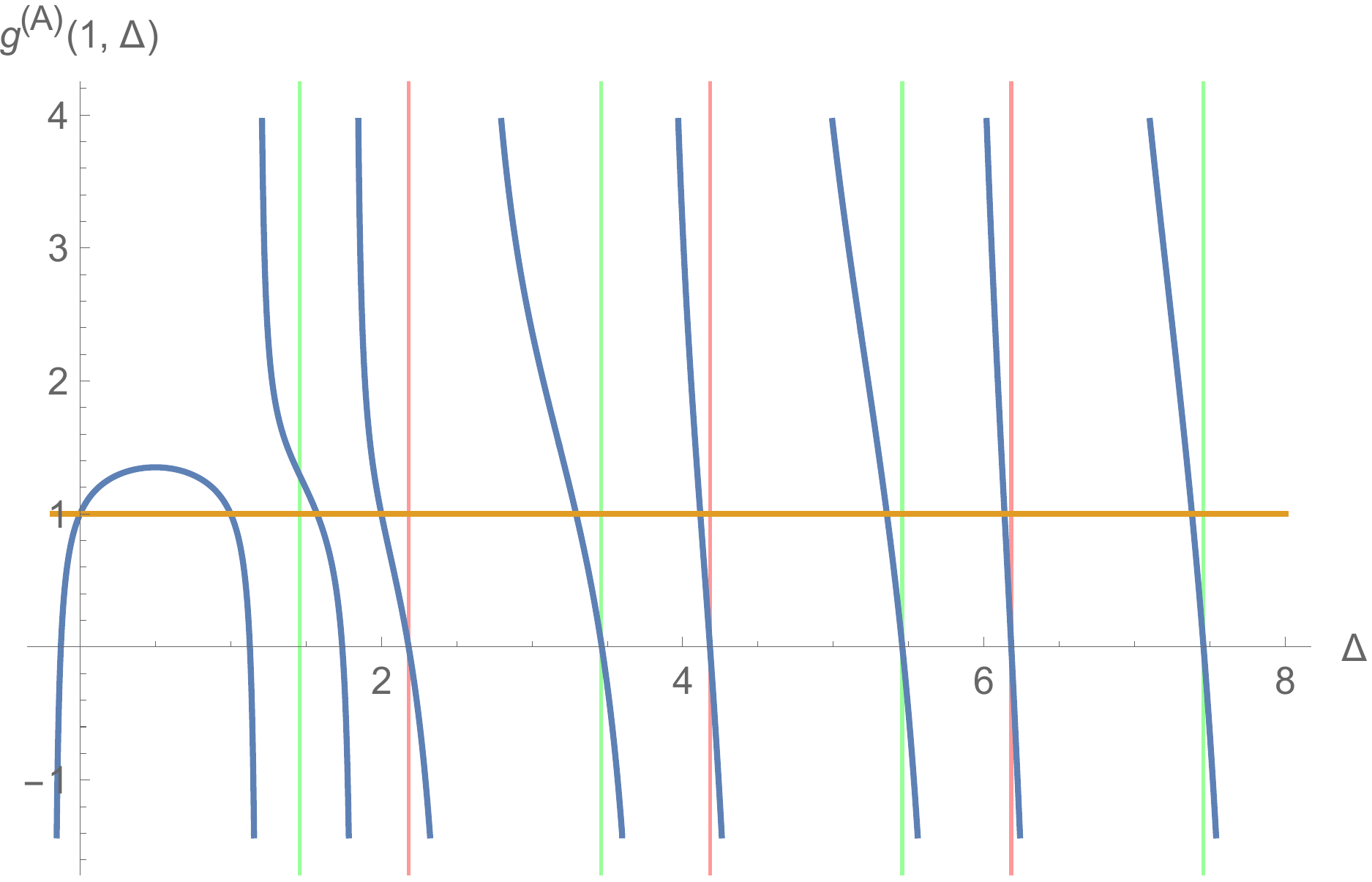}
\caption{The spectrum of scalar type A/C bilinears in 1d. Red vertical lines are asymptotes corresponding to $-2\Delta_\phi+2n$ and green vertical lines are asymptotes corresponding to $-2\Delta_\chi+2n$. 
\label{1deigen}
}
\end{center}
\end{figure}

Solving for the scaling dimensions of type A/C bilinears in $d=1$, we find that the low-lying eigenvalues are 
\begin{equation}
\Delta=1,\; 1.57, \; 2, \; 3.29, \; 4.12, \; 5.36, \; 6.14, \; 7.38,\;
8.15, \; 9.39, \; 10.15, \; 11.40, \ldots
\end{equation} 
The plot for the eigenvalues is shown in figure \ref{1deigen}.

The smallest positive eigenvalue, $\Delta=1$, is the continuation of the solution $\Delta=d$ present for any $d$. As discussed  in section (\ref{threeeps}), it may correspond to
a redundant operator. The next scaling dimension, $\Delta=1.57317$, may correspond to a mixture involving $\phi^{abc} \phi^{abc}$.
The appearance of scaling dimension $2$, which was also seen for the fermionic SYK and tensor models, means that the its dual\footnote{Of course, as observed in \cite{Choudhury:2017tax,Bulycheva:2017ilt}, there are important differences between the holographic duals of tensor models and SYK models.} should involve dilaton 
gravity in AdS$_2$ \cite{Almheiri:2014cka,Maldacena:2016upp,Engelsoy:2016xyb,Jensen:2016pah}.

Let us also list the type B scaling dimensions, i.e. the ones corresponding to operators $\phi^{abc} \partial_t^{2n} \chi^{abc}$. 
Here we find real solutions $\Delta= 1.01,~2.96,~4.94,~6.93,~ \ldots$.

For large excitation numbers $n$, the type A/C scaling 
dimensions appear to (slowly) approach $-2\Delta_\phi +2n$ and $-2\Delta_\chi+2n$ rather than $2\Delta_\phi +2n$ and $2\Delta_\chi+2n$, as shown in figure \ref{1d}. 
The type B scaling dimensions also appear to slowly approach  $-\Delta_\phi - \Delta_\chi + 2n$ rather than $\Delta_\phi + \Delta_\chi + 2n$. This is likely due to the fact that $\Delta_\phi$ is negative. Further work is needed to understand better the new features of the large $N$ solution in the regime where $d< 1.35$ and $\Delta_\phi < 0$.

\section{Discussion}

In this paper we presented exact results for the $O(N)^3$ invariant theory (\ref{prism}) in the prismatic large $N$ limit where $g_1 N^3$ is held fixed.
This approach may be generalized to an $O(N)^p$ invariant theory of a rank-$p$ bosonic tensor $\phi^{a_1 \ldots a_p}$, with odd $p\geq 3$. It has a positive potential of
order $2p$:
\begin{equation}
S_{2p}= \int d^d x \left ( \frac {1} {2} (\partial_\mu \phi^{abc})^2 + 
{g_1\over (2p)!} (\phi^p)^{a_1 \ldots a_p}   (\phi^p)^{a_1 \ldots a_p} 
\right )
\ .
\label{prismnew}
\end{equation}
To solve these models in the large $N$ limit where $g_1 N^p$ is held fixed,
we may rewrite the action with the help of an additional tensor field $\chi$:
\begin{equation}
S= \int d^d x \left ( \frac {1} {2} (\partial_\mu \phi^{abc})^2 + 
{g\over p!} (\phi^p)^{a_1 \ldots a_p}   \chi^{a_1 \ldots a_p}   -\\\frac{1}{2}\chi^{a_1 \ldots a_p}\chi^{a_1 \ldots a_p}\right )\ .
\label{prismauxp}
\end{equation}
For discussions of the structure of the interaction vertex with odd $p>3$, see  
\cite{Klebanov:2016xxf,Ferrari:2017jgw,Gubser:2018yec}.
The models (\ref{prismnew}) are tensor counter-parts of the SYK-like models introduced in 
\cite{Murugan:2017eto}; therefore, the Schwinger-Dyson equations derived there should be applicable to the tensor models.
It would be interesting to study the large $N$ solution of theories with $p>3$ in more detail using methods analogous to the ones used for $p=3$. 

In this paper we analyzed the renormalization of the prismatic theory at the two-loop order, using the beta functions in  \cite{Gracey:2015fia,Osborn:2017ucf}. 
The general four-loop terms are also given there, and it would be interesting to study the effects they produce. 
It should be possible to extend the calculations to even higher loops by modifying the calculations in \cite{Hager:2002uq} to an arbitrary tensorial interaction, which we leave as a possible avenue for future work. In this context, it would also be interesting to study the possibility of fixed points with other large $N$ scalings, perhaps dominated by the ``wheel" interaction ($g_2$) of figure \ref{8inter}, in addition to the large $N$ fixed point dominated by the prism interaction ($g_1$) studied in this paper.\footnote{A $d=0$ theory with wheel interactions was studied
in \cite{Lionni:2017xvn}.} 

Another interesting extension of the $O(N)^3$ symmetric model (\ref{prism}) is to add a 2-component Majorana fermion $\psi^{abc}$, so that the fields can be assembled into a $d=3$
${\cal N}=1$ superfield
\begin{equation}
\Phi^{abc} = \phi^{abc} + \bar \theta \psi^{abc} +  \bar \theta \theta \chi^{abc}
\end{equation}
Then the prismatic scalar potential follows if we assume a tetrahedral superpotential for $\Phi^{abc}$ \cite{Klebanov:2016xxf}. 
Large $N$ treatments of supersymmetric tensor and SYK-like models with two supercharges have been given in \cite{Murugan:2017eto,Chang:2018sve}, and we expect the solution of the
${\cal N}=1$ super-tensor model in $d<3$ to work analogously. An advantage of the tensor QFT approach is that one can also develop the $3-\epsilon$ expansion
using the standard renormalized perturbation theory. In the supersymmetric case, it is sufficient to introduce only three coupling constants:
\begin{align}
 W& = g_1 \Phi^{a_1 b_1 c_1} \Phi^{a_1 b_2 c_2} \Phi^{a_2 b_1 c_2} \Phi^{a_2 b_2 c_1}\notag \\
& + g_2 \big(\Phi^{a_{1}b_{1}c_{1}}\Phi^{a_{1}b_{1}c_{2}}\Phi^{a_{2}b_{2}c_{1}}\Phi^{a_{2}b_{2}c_{2}} +
\Phi^{a_{1}b_{1}c_{1}}\Phi^{a_{2}b_{1}c_{1}}\Phi^{a_{1}b_{2}c_{2}}\Phi^{a_{2}b_{2}c_{2}} 
+\Phi^{a_{1}b_{1}c_{1}}\Phi^{a_{1}b_{2}c_{1}}\Phi^{a_{2}b_{1}c_{2}}\Phi^{a_{2}b_{2}c_{2}}\big) \notag \\
& + g_3  \Phi^{a_{1}b_{1}c_{1}}\Phi^{a_{1}b_{1}c_{1}}\Phi^{a_{2}b_{2}c_{2}}\Phi^{a_{2}b_{2}c_{2}}\ ,
\end{align}
and it is possible to find explicit expressions for the beta functions and operator scaling dimensions \cite{PT:2018}.
Also, directly in $d=3$ it is possible to couple the ${\cal N}=1$ theory with the above superpotential to $O(N)_{k_1} \times O(N)_{k_2}\times O(N)_{k_3}$ 
supersymmetric Chern-Simons gauge theory with levels $k_1, k_2, k_3$,
and derive the corresponding beta functions for couplings $g_i$ \cite{PT:2018}.

\section*{Acknowledgments}

We are grateful to C.-M. Chang, M. Rangamani, D. Stanford, E. Witten and J. Yoon for useful discussions.
IRK thanks the Yukawa Institute for Theoretical Physics where some of his work on this paper was carried out
during the workshop YITP-T-18-04 ``New Frontiers in String Theory 2018". SP thanks the Princeton Center for Theoretical Science for hospitality as well as the International Centre for Theoretical Sciences, Bengaluru where some of his work on this paper was carried out during the program - ``AdS/CFT at 20 and Beyond" (ICTS/adscft20/2018/05).
The work of SG was supported in part by the US NSF under Grant No.~PHY-1620542.
The work of IRK and FP was supported in part by the US NSF under Grant No.~PHY-1620059. The work of SP was supported in part by a DST-SERB Early Career Research Award (ECR/2017/001023) and a DST INSPIRE Faculty Award.  The work of GT was supported in part by  the MURI grant W911NF-14-1-0003 from ARO and by DOE grant de-sc0007870.

\appendix
\section{The two-loop beta functions and anomalous dimensions}

In this Appendix we state our explicit two-loop results for the $O(N)^3$ invariant theory with the 8 coupling constants and interaction terms
\begin{align}
&{g_1\over 6!} \phi^{a_1 b_1 c_1} \phi^{a_1 b_2 c_2} \phi^{a_2 b_1 c_2} \phi^{a_3 b_3 c_1} \phi^{a_3 b_2 c_3} \phi^{a_2 b_3 c_3} +
{g_2\over 6!}
\phi^{a_{1}b_{1}c_{1}}\phi^{a_{1}b_{2}c_{2}}\phi^{a_{2}b_{2}c_{3}}\phi^{a_{2}b_{3}c_{1}}\phi^{a_{3}b_{3} 
  c_{2}}\phi^{a_{3}b_{1}c_{3}}\notag \\
&+ {g_3\over 3\cdot 6!}\big (
\phi^{a_1 b_1 c_1}\phi^{a_2 b_1 c_1} \phi^{a_1 b_2 c_2}\phi^{a_2 b_2 c_3} \phi^{a_3 b_3 c_2}\phi^{a_3 b_3 c_3} +
\phi^{a_1 b_1 c_1}\phi^{a_1 b_2 c_1} \phi^{a_2 b_1 c_2}\phi^{a_2 b_2 c_3} \phi^{a_3 b_3 c_2}\phi^{a_3 b_3 c_3}\notag \\
& +
\phi^{a_1 b_1 c_1}\phi^{a_2 b_1 c_1} \phi^{a_1 b_2 c_2}\phi^{a_2 b_3 c_2} \phi^{a_3 b_2 c_3}\phi^{a_3 b_3 c_3} \big ) 
\notag \\
&+ {g_4\over 3\cdot 6!}\big (
\phi^{a_1 b_1 c_1}\phi^{a_1 b_1 c_2} \phi^{a_2 b_2 c_2}\phi^{a_2 b_2 c_3} \phi^{a_3 b_3 c_3}\phi^{a_3 b_3 c_1} +
\phi^{a_1 b_1 c_1}\phi^{a_2 b_1 c_1} \phi^{a_2 b_2 c_2}\phi^{a_3 b_2 c_2} \phi^{a_3 b_3 c_3}\phi^{a_1 b_3 c_3}\notag \\
& +
\phi^{a_1 b_1 c_1}\phi^{a_1 b_2 c_1} \phi^{a_2 b_2 c_2}\phi^{a_2 b_3 c_2} \phi^{a_3 b_3 c_3}\phi^{a_3 b_1 c_3} \big ) 
\notag \\
&+ {g_5\over 3\cdot 6!}\big (
\phi^{a_1 b_1 c_1}\phi^{a_1 b_2 c_2} \phi^{a_2 b_1 c_2}\phi^{a_3 b_2 c_1} \phi^{a_2 b_3 c_3}\phi^{a_3 b_3 c_3} +
\phi^{a_1 b_1 c_1}\phi^{a_2 b_1 c_2} \phi^{a_1 b_2 c_2}\phi^{a_1 b_2 c_3} \phi^{a_3 b_2 c_3}\phi^{a_3 b_3 c_3}\notag \\
& +
\phi^{a_1 b_1 c_1}\phi^{a_2 b_2 c_1} \phi^{a_2 b_1 c_2}\phi^{a_1 b_2 c_3} \phi^{a_3 b_3 c_2}\phi^{a_3 b_3 c_3} \big ) 
\notag \\
&+ {g_6\over 6!} \phi^{abc} \phi^{abc} \phi^{a_1 b_1 c_1} \phi^{a_1 b_2 c_2} \phi^{a_2 b_1 c_2} \phi^{a_2 b_2 c_1}  \notag\\
&+ {g_7\over 3\cdot 6!} \phi^{abc} \phi^{abc} (
\phi^{a_1 b_1 c_1} \phi^{a_1 b_1 c_2} \phi^{a_2 b_2 c_1} \phi^{a_2 b_2 c_2} + 
\phi^{a_1 b_1 c_1} \phi^{a_2 b_1 c_1} \phi^{a_1 b_2 c_2} \phi^{a_2 b_2 c_2} + 
\phi^{a_1 b_1 c_1} \phi^{a_1 b_2 c_1} \phi^{a_2 b_1 c_2} \phi^{a_2 b_2 c_2} )\notag \\  
&+ {g_8\over 6!} (\phi^{abc} \phi^{abc})^3
\ .
\label{allinter}
\end{align}
We find
\begin{align}\label{beta1}
\beta_{1}=&-2 g_1 \epsilon+\frac{1}{270 (8\pi) ^2}\Big((g_5^2+3 (g_1^2+8 g_6^2)) N^3+3 (3 g_5^2+4 (2 g_1+3 g_2+4 g_6) g_5+6 g_1 (g_1+3 g_2)) N^2\notag\\
&+2 (32 g_5^2+(90 g_1+72 g_2+96 g_6) g_5+6 g_4 (9 g_2+4 g_5)+9 (5 g_1^2+6 g_2 g_1+16 g_6 g_1+8 g_7 g_1+9 g_2^2\notag\\
&+24 g_2 g_6)) N+2 g_3^2 (N (N+6)+55)+2 g_3 (9 N (g_1 (N+8)+8 g_2)+6 g_4 (N+6)\notag\\
&+2 g_5 (N+10) (2 N+5)+2 (60 g_1+63 g_2+96 g_6+16 g_7))+2 (36 g_4^2+36 (5 g_1+3 g_2+2 g_5) g_4\notag\\
&+80 g_5^2+4 g_5 (45 g_1+4 (9 g_2+6 g_6+8 g_7))+3 (34 g_1^2+12 (7 g_2+4 g_6+2 g_7+20 g_8) g_1+27 g_2^2\notag\\
&+128 g_6^2+48 g_2 (g_6+2 g_7))\Big)
\end{align}

\begin{align}
\beta_{2}=&-2 g_2 \epsilon+\frac{1}{270 (8\pi) ^2}\Big(g_5 (12 g_1+g_5) N^2+2 (13 g_5^2+18 (g_1+g_2) g_5+9 g_1 (g_1+2 g_4+8 g_6)+72 g_2 g_7) N\notag\\
&+2 g_3^2 (N (N+6)+19)+2 g_3 (3 N (3 g_2 (N+4)+8 g_1)+6 g_4 (N+2)+6 g_5 (N+6)+30 g_1+36 g_2\notag\\
&+32 g_7)+2 (36 g_1^2+54 g_2 g_1+96 g_7 g_1+45 g_2^2+12 g_4^2+20 g_5^2+12 g_4 (3 g_1+9 g_2+2 g_5)\notag\\
&+12 g_5 (4 g_1+3 g_2+8 g_6)+72 g_2 g_7+720 g_2 g_8)\Big)
\end{align}

\begin{align}
\beta_{3}=&-2 g_3 \epsilon+\frac{1}{270 (8\pi) ^2}\Big(2 (g_5^2+8 g_7^2) N^3+3 (6 g_1^2+12 g_5 g_1+27 g_2^2+5 g_5^2) N^2+2 (83 g_5^2+2 (66 g_1\notag\\
&+63 g_2+48 g_6+64 g_7) g_5+9 (2 g_1+3 g_2) (4 g_1+3 g_2)+96 (g_1+3 g_2) g_7) N+g_3^2 (N (N (2 N+31)\notag\\
&+244)+388)+18 g_4^2 (N (N+16)+12)+12 g_4 (3 g_1 (N+1) (N+14)+g_5 (5 N (N+6)+72)\notag\\
&+(N+2) (9 g_2 (N+3)+8 g_7 N)+96 g_6+64 g_7)+4 g_3 (3 g_4 (N (N (N+6)+28)+102)\notag\\
&+N (g_5 (11 N+74)+6 (g_1+3 g_2+4 g_7) N+66 g_1+72 g_2+60 g_6+84 g_7)+194 g_5\notag\\
&+3 (71 g_1+81 g_2+32 g_6+76 g_7+120 g_8))+4 (92 g_5^2+2 (93 g_1+90 g_2+72 g_6+80 g_7) g_5\notag\\
&+128 g_7^2+9 (7 g_1^2+15 g_2 g_1+9 g_2^2+24 (g_1+g_2) g_6)+144 (g_1+g_2) g_7)\Big)
\end{align}

\begin{align}
\beta_{4}=&-2 g_4 \epsilon+\frac{1}{270 (8\pi) ^2}\Big((g_5^2+8 g_7^2) N^3+4 g_5 (3 g_1+g_5) N^2+6 (3 g_1^2+9 g_5^2+8 (g_1+3 g_2) g_7\notag\\
&+2 g_5 (5 g_1+9 g_2+4 (g_6+3 g_7))) N+2 g_3^2 (N (N (N+7)+34)+113)+9 g_4^2 (N (N+2)^2+52)\notag\\
&+4 g_3 (9 g_2 (N+2)^2+3 g_1 (N+1) (N+13)+N (g_4 (6 N+75)+g_5 (6 N+31)+8 g_7 (N+4))\notag\\
&+16 (3 g_4+5 g_5+6 g_6+5 g_7))+12 g_4 (3 g_1 (N+12)+2 g_5 (N (N+6)+13)+8 N (g_7 (N+2)+3 g_6)\notag\\
&+48 g_2+44 g_7+120 g_8)+2 (54 g_1^2+162 g_2 g_1+96 g_7 g_1+81 g_2^2+58 g_5^2+128 g_7^2\notag\\
&+4 g_5 (27 g_1+27 g_2+24 g_6+32 g_7))\Big)
\end{align}

\begin{align}
\beta_{5}=&-2 g_5 \epsilon +\frac{2}{270 (8\pi) ^2}\Big( (3 (g_1 g_5+8 g_6 g_7) N^3+2 (9 g_1^2+9 (3 g_2+g_5) g_1+g_5 (27 g_2+6 g_5+16 g_7)) N^2\notag\\
&+(99 g_1^2+6 (45 g_2+35 g_5+36 g_6+40 g_7) g_1+81 g_2^2+216 g_2 (g_5+g_6)\notag\\
&+4 g_5 (21 g_5+42 g_6+38 g_7)) N+g_3^2 (N (5 N+52)+161)+36 g_4^2 (N+3)\notag\\
&+3 g_4 (12 g_1 (N (N+5)+12)+g_5 (N (N (N+6)+52)+132)+6 (N+2) (4 g_6 N+9 g_2)\notag\\
&+96 g_6+64 g_7)+2 g_3 (6 g_1 (N (3 N+16)+37)+9 g_2 (10 N+23)+N (g_4 (6 N+39)\notag\\
&+g_5 (N (N+13)+97)+24 g_6 (N+4))+6 (23 g_4+33 g_5+32 g_6+24 g_7))+270 g_1^2\notag\\
&+243 g_2^2+212 g_5^2+432 g_1 g_2+444 g_1 g_5+504 g_2 g_5+432 g_1 g_6+432 g_2 g_6\notag\\
&+384 g_5 g_6+384 g_1 g_7+288 g_2 g_7+328 g_5 g_7+768 g_6 g_7+720 g_5 g_8)\Big)
\end{align}

\begin{align}
\beta_{6}=&-2 g_6 \epsilon+\frac{2}{270 (8\pi) ^2}\Big(2 (g_5 g_7+3 g_6 (g_1+12 g_8)) N^3+(6 (9 g_2 g_6+4 (g_1+2 g_6) g_7)\notag\\
&+g_5 (3 g_1+12 g_6+10 g_7+72 g_8)) N^2+(7 g_5^2+2 (3 g_1+9 g_2+12 g_6+32 g_7+72 g_8) g_5\notag\\
&+3 (3 g_1+12 (2 g_6+g_7+12 g_8) g_1+48 g_6^2+8 (3 g_2+5 g_6) g_7)) N+g_3^2 (4 N+6)\notag\\
&+3 g_4 (12 g_1 N+g_5 (N (N+6)+10)+4 g_7 (N+2)+18 g_2+60 g_6)+2 g_3 (6 g_1 (N+4)\notag\\
&+g_5 (N (N+6)+19)+3 g_6 (N (N+10)+4)+2 g_7 N (N+5)+9 g_2+21 g_4+18 (g_7+4 g_8))\notag\\
&+13 g_5^2+48 g_7^2+36 g_1 g_2+30 g_1 g_5+18 g_2 g_5+48 g_1 g_6+72 g_2 g_6+108 g_5 g_6+120 g_1 g_7\notag\\
&+36 g_2 g_7+92 g_5 g_7+120 g_6 g_7+432 g_2 g_8+144 g_5 g_8+1296 g_6 g_8\Big)
\end{align}

\begin{align}
\beta_{7}=&-2 g_7 \epsilon+\frac{1}{270 (8\pi) ^2}\Big(4 (3 g_5 g_6+g_7 (2 g_3+3 g_4+36 g_8)) N^3+(10 g_3^2+24 (g_4+3 g_6\notag\\
&+2 (g_7+6 g_8)) g_3+7 g_5^2+3 (9 g_4^2+8 (3 g_6+2 (g_7+9 g_8)) g_4+8 (5 g_7^2+(g_1+3 g_2+2 g_5) g_7\notag\\
&+(3 g_1+g_5) g_6))) N^2+(9 g_1^2+54 g_5 g_1+72 g_6 g_1+216 g_7 g_1+48 g_3^2+63 g_4^2+22 g_5^2+216 g_6^2\notag\\
&+216 g_7^2+216 g_2 g_6+216 g_5 g_6+144 g_2 g_7+160 g_5 g_7+576 g_6 g_7+144 (3 g_1+9 g_2+5 g_5) g_8\notag\\
&+6 g_4 (6 g_1+18 g_2+21 g_5+36 g_6+52 g_7+72 g_8)+4 g_3 (3 g_1+9 g_2+36 g_4+19 g_5+42 g_6\notag\\
&+90 g_7+144 g_8)) N+2 (27 g_1^2+3 (9 g_2+23 g_3+30 g_4+12 g_5+48 g_6+40 g_7+144 g_8) g_1\notag\\
&+9 g_2 (7 g_3+6 (g_5+2 g_6+4 g_7))+2 (31 g_3^2+(81 g_4+50 g_5+114 g_6+112 g_7+216 g_8) g_3\notag\\
&+54 g_4^2+21 g_5^2+108 g_6^2+96 g_7^2+66 g_5 g_6+106 g_5 g_7+144 g_6 g_7+72 (2 g_5+9 g_7) g_8\notag\\
&+3 g_4 (17 g_5+36 g_6+66 g_7+72 g_8)))\Big)
\end{align}

\begin{align}\label{beta8}
\beta_{8}=&-2 g_8 \epsilon+\frac{1}{270 (8\pi) ^2}
\Big(2 (g_5 (2 (3 g_6 (N^2+N+3)+7 g_7 (N+1)+36 g_8)+3 g_1)+2 (3 g_6^2 N^3\notag\\
&+g_7^2 N^3+18 g_8^2 (3 N^3+22)+3 g_7^2 N^2+12 g_6 g_7 N^2+72 g_8 (g_7 (N^2+N+1)+3 g_6 N)+9 g_6^2 N\notag\\
&+21 g_7^2 N+12 g_6 g_7 N+g_1 (9 g_6 N+6 g_7)+6 g_6^2+23 g_7^2+9 g_2 g_6+48 g_6 g_7)+g_5^2 (N+1)\notag\\
&+3 g_4 (2 (6 g_6 N+g_7 (N (N+3)+5)+36 g_8)+3 g_5))+g_3^2 (2 N+9)+4 g_3 (3 g_4 N+3 g_6 (2 N+5)\notag\\
&+2 g_7 (N (N+3)+7)+36 g_8 N+2 g_5)+9 g_2^2+39 g_4^2\Big)
\end{align}
and 
\begin{align}
\gamma_\phi =& \frac{1}{12\cdot 90^2 (8 \pi )^4}\Big((3 g_1^2+9 g_2^2+g_3^2+3 g_4^2+g_5^2+12 g_6^2+4 g_7^2+72 g_8^2) N^6+(6 g_3^2+2 (3 g_1+9 g_2\notag\\
&+6 (g_4+g_5)+8 g_7) g_3+9 g_4^2+5 g_5^2+12 g_7^2+54 g_1 g_2+24 g_1 g_5+24 g_5 g_6+48 g_6 g_7\notag\\
&+12 g_4 (g_5+2 g_7)+144 g_7 g_8) N^5+(81 g_1^2+12 (9 g_3+6 g_4+5 g_5+12 g_6+2 g_7) g_1+81 g_2^2\notag\\
&+39 g_3^2+27 g_4^2+51 g_5^2+36 g_6^2+84 g_7^2+108 g_3 g_4+76 g_3 g_5+72 g_4 g_5+96 g_3 g_6+144 g_4 g_6\notag\\
&+48 g_5 g_6+80 g_3 g_7+96 g_4 g_7+88 g_5 g_7+48 g_6 g_7+36 g_2 (2 g_3+g_4+4 g_5+2 g_7)\notag\\
&+144 (g_3+g_4+3 g_6+g_7) g_8) N^4+(102 g_1^2+6 (75 g_2+47 g_3+54 g_4+64 g_5+24 g_6\notag\\
&+68 g_7+24 g_8) g_1+54 g_2^2+160 g_3^2+171 g_4^2+143 g_5^2+120 g_6^2+148 g_7^2+432 g_8^2+288 g_3 g_4\notag\\
&+344 g_3 g_5+336 g_4 g_5+336 g_3 g_6+288 g_4 g_6+360 g_5 g_6+336 g_3 g_7+336 g_4 g_7+296 g_5 g_7\notag\\
&+336 g_6 g_7+144 (2 g_3+3 (g_4+g_5)+g_7) g_8+18 g_2 (19 g_3+24 g_4+14 g_5+32 g_6+12 g_7+24 g_8)) N^3\notag\\
&+2 (189 g_1^2+6 (45 g_2+58 g_3+66 g_4+49 g_5+72 g_6+54 g_7+108 g_8) g_1+216 g_2^2+177 g_3^2\notag\\
&+189 g_4^2+176 g_5^2+216 g_6^2+120 g_7^2+318 g_3 g_4+330 g_3 g_5+336 g_4 g_5+360 g_3 g_6+288 g_4 g_6\notag\\
&+312 g_5 g_6+328 g_3 g_7+312 g_4 g_7+372 g_5 g_7+336 g_6 g_7+72 (4 g_3+4 g_4+5 g_5+4 g_7) g_8\notag\\
&+18 g_2 (17 g_3+19 g_4+20 g_5+12 g_6+26 g_7+12 g_8)) N^2+4 (81 g_1^2+3 (63 g_2+63 g_3\notag\\
&+51 g_4+64 g_5+60 g_6+70 g_7+36 g_8) g_1+81 g_2^2+87 g_3^2+72 g_4^2+90 g_5^2+72 g_6^2+96 g_7^2\notag\\
&+207 g_3 g_4+185 g_3 g_5+189 g_4 g_5+156 g_3 g_6+216 g_4 g_6+204 g_5 g_6+184 g_3 g_7+174 g_4 g_7\notag\\
&+182 g_5 g_7+168 g_6 g_7+36 (6 g_3+3 g_4+5 g_5+12 g_6+4 g_7) g_8+9 g_2 (23 g_3+18 g_4+19 g_5\notag\\
&+24 g_6+18 g_7+36 g_8)) N+4 (48 g_1^2+(90 g_2+78 g_3+90 g_4+84 g_5+72 g_6+60 g_7+72 g_8) g_1\notag\\
&+45 g_2^2+43 g_3^2+51 g_4^2+42 g_5^2+48 g_6^2+52 g_7^2+144 g_8^2+72 g_3 g_4+82 g_3 g_5+78 g_4 g_5+96 g_3 g_6\notag\\
&+72 g_4 g_6+72 g_5 g_6+84 g_3 g_7+96 g_4 g_7+76 g_5 g_7+96 g_6 g_7+18 g_2 (4 g_3+5 g_4+5 g_5+4 (g_6+g_7))\notag\\
&+72 (g_3+2 g_4+g_5+2 g_7) g_8)\Big)
\label{gammaphi}
\end{align}
At the two-loop level we also find the relation $\gamma_{\phi^2}= 32 \gamma_\phi$.

We can study the anomalous dimensions for quartic operators
\begin{gather}
O_1 = O_{\rm tetra} = \phi^{a_1 b_1 c_1}\phi^{a_1 b_2 c_2} \phi^{a_2 b_1 c_2} \phi^{a_2 b_2 c_1},\notag\\ O_2=O_{\rm pillow} = \frac13\left(\phi_{a_1 b_1 c_1}\phi_{a_2 b_1 c_1}\phi_{a_1 b_2 c_2}\phi_{a_2 b_2 c_2}+\phi_{a_1 b_1 c_1}\phi_{a_1 b_2 c_1}\phi_{a_2 b_1 c_2}\phi_{a_2 b_2 c_2}+\phi_{a_1 b_1 c_1}\phi_{a_1 b_1 c_2}\phi_{a_2 b_2 c_1}\phi_{a_2 b_2 c_2}\right)\notag\\
O_3 = O_{\rm d.t.} = \phi^{a_1 b_1 c_1}\phi^{a_1 b_1 c_1}\phi^{a_2 b_2 c_2}\phi^{a_2 b_2 c_2}.
\end{gather}
The matrix of anomalous dimensions for quartic operators can be written in the following way
\begin{gather}
\gamma^{11}_{O} = \frac{1}{720\pi^2} \left(2 (6 g_1 + 2 g_3 + 3 g_4+ 5 g_5+2 g_7+12 g_8) + g_1(N^3+12 N +8) + 4 (g_5 + 3 g_6 + g_7) N +\right.\notag\\\left.+ 9 g_2 N^2 + 2 g_5 N^2 + g_3 \left(6 N +N^2\right)\right), \notag\\
\gamma^{12}_{O} = \frac{1}{2160\pi^2} \left(2 (9 g_2 + 9 g_3 + 6 g_4 + 11 g_5 + 12 g_6+8 g_7) + 6 g_1 (6 + 3 N + 2N^2) + 36 g_2 N + 6 g_4 N + \right.\notag\\\left.+ 12 g_6 (2 N + N^2) + 2 g_3 (5 N + N^2) + g_5 (24 N + 5 N^2 + N^3)\right)\notag\\
\gamma^{13}_{O} = \frac{1}{180\pi^2} \left( 6 g_2 + 2 g_3 + 6 g_1 N + g_6 (8+N^3) + g_5 (2 + 2 N + N^2)\right)\notag\\
\gamma^{21}_{O} = \frac{1}{720\pi^2} \left( 2 (12 g_1 + 9 g_2 + 11 g_3 + 12 g_4 + 9 g_5+ 12 g_6 + 8 g_7) + g_5 N^3 +\right.\notag\\\left.+ 2 (3 g_1 + 9 g_2 + 7 g_3 + 9 g_4 + 9 g_5 + 6 g_6 + 10 g_7) N + 2 (3 (g_1 + g_3 + g_4) + g_5) N^2\right)\notag\\
\gamma^{22}_{O} = \frac{1}{2160 \pi^2} \left( 64 g_3 + 66 g_4 + 62 g_5 + 48 g_6 + 60 g_7 + 72 g_8 + 6 g_1 (N+1)(N+8) + 18 g_2 (4 + 2N + N^2) +\right.\notag\\ \left. +  3 g_4 (18 N + 4 N^2 + N^3) + 2 g_3 (27 N + 6 N^2 + N^3) + 4 (6 g_6 N + 4 g_7 (2 N + N^2) + g_5 (10 N + 3 N^2)) \right)\notag\\
\gamma^{23}_{O} = \frac{1}{180\pi^2} \left(6 g_3 + 6 g_4 + 4 g_5 + 8 g_7 + 3 g_1 (N+2) + 9 g_2 N+ 5 g_5 N + g_7 N^3+ 3 g_4 (N^2+ N) +\right.\notag\\\left.+ 2 g_3 (2 N + N^2)\right)\notag\\
\gamma^{31}_{O} = \frac{1}{720 \pi^2} \left(3 g_2 + 3 g_5 +4 g_6 + 8 g_7 + 3 g_1 N + g_3 (5 + 2N) + 6 g_4 N + g_5 (N^2+ N) +\right.\notag\\\left.+ 4(g_7 N + 9 g_8 N + g_7 N^2) + 2 g_6 (3 N+ N^3)\right)\notag\\
\gamma^{32}_{O} = \frac{1}{2160\pi^2} \left(6 g_1 + 7 g_5 + 24 g_6 + 22 g_7+ 36 g_8 + 2 g_3 (5 +3 N + N^2) + 3 g_4 (5 + 3N+N^2) + 7 g_5 N +\right.\notag\\\left.+ 12 g_6 (N + N^2) + 36 g_8 (N + N^2) + 2 g_7 (13 N + 3 N^2 + N^3)\right) 
\end{gather}
The results for the quartic operator dimensions in the prismatic large $N$ limit are listed in (\ref{tetrapillow}).

A consistent truncation of the system of eight coupling constants is to keep only $g_8$ non-vanishing, since the triple-trace term is the only one which has $O(N^3)$ symmetry.
Then we find 
\begin{align}
\beta_{8}=-2 g_8 \epsilon+\frac{1}{15 (8\pi) ^2}
g_8^2 (3 N^3+22)\ , \qquad \gamma_\phi =& \frac{1}{1350 (8 \pi )^4} g_8^2 (N^3 + 2)(N^3+4)\ ,
\end{align}
in agreement with \cite{Hager:2002uq, Osborn:2017ucf}.
Thus, there is a fixed point with 
\begin{align}
g_8^* = \frac{30 (8\pi) ^2 \epsilon}{3 N^3+22}\ , \qquad g_i^*=0, \quad i=1, \ldots, 7\ . 
\end{align}
At this fixed point,
\begin{align}
\partial \beta_8/\partial g_8 = -2\epsilon+ \frac{2}{15 (8\pi) ^2}
g_8^* (3 N^3+22)= 2\epsilon + \mathcal{O}(\epsilon^2)\ ,
\end{align}
 so the triple-trace operator is irrelevant. However, the other 7 operators appear to be relevant for
sufficiently large $N$. For example,
\begin{align}
{\partial \beta_1\over \partial g_1}=- 2 \epsilon+ {2 g_8^*\over 9 (8\pi)^2}= \epsilon \left (-2 + \frac {20}{ 3 (3N^3+ 22)} \right ) + \mathcal{O}(\epsilon^2) \ .
\end{align}
So, this fixed point has 7 unstable directions.
Examination of 4-loop and higher corrections \cite{Hager:2002uq, Osborn:2017ucf}
shows that the $3-\epsilon$ expansions of operator dimensions at this fixed point do not generally have a finite large $N$ limit starting with order $\epsilon^3$.
This is in contrast with the prismatic fixed point where all the $g_i^*$ are non-vanishing and scale as (\ref{scalings}); as a result, the large $N$ limit is smooth.

\bibliographystyle{ssg}
\bibliography{phi6_tensor_model}

\end{document}